\documentclass[journal,10pt,compsoc]{IEEEtran}

\usepackage{amsmath}
\usepackage{amsthm}
\usepackage{bm}
\usepackage[overload]{empheq}
\usepackage[noend]{algpseudocode}
\usepackage{algorithm}
\makeatletter
\def\BState{\State\hskip-\ALG@thistlm}
\makeatother

\ifCLASSOPTIONcompsoc
  \usepackage[nocompress]{cite}
\else
  \usepackage{cite}
\fi
\usepackage[pdftex]{graphicx}

\ifCLASSINFOpdf
\else
\fi
\usepackage{enumitem}
\usepackage{xcolor}
\usepackage{amssymb}

\usepackage{hyperref}

\definecolor{navy}{rgb}{.0,.0,.0}

\usepackage{tikz}

\renewcommand{\v}[1]{\mathbf{#1}}
\newcommand{\vg}[1]{\bm{#1}}
\newcommand{\tran}[0]{\mathsf{T}}

\newcommand\blfootnote[1]{%
  \begingroup
  \renewcommand\thefootnote{}\footnote{#1}%
  \addtocounter{footnote}{-1}%
  \endgroup
}

\usepackage{tabularx}

\begin{document}
%
\title{Online Resource Inference in Network Utility Maximization Problems}
%
%
%
%

\author{Stefano~D'Aronco,
        Pascal~Frossard} 
\IEEEtitleabstractindextext{%
\begin{abstract}
The amount of transmitted data in computer networks is expected to grow considerably in the future, putting  more and more pressure on the network infrastructures.
In order to guarantee a good service, it then becomes fundamental to use the network resources  efficiently. Network Utility Maximization (NUM) provides a framework to optimize the rate allocation when network resources are limited. Unfortunately, in the scenario where the amount of available resources is not known a priori, classical NUM solving methods do not offer a viable solution. To overcome this limitation we design an overlay rate allocation scheme that attempts to infer the actual amount of available network resources while coordinating the users rate allocation.
Due to the general and complex model assumed for the congestion measurements, a passive learning of the available resources would not lead to satisfying performance. The coordination scheme must then perform active learning in order to speed up the resources estimation and quickly increase the system performance. 
By adopting an optimal learning formulation we are able to balance the tradeoff between an accurate estimation, and an effective resources exploitation in order to maximize the long term quality of the service delivered to the users. 
\end{abstract}

}

\maketitle

\IEEEdisplaynontitleabstractindextext

%
\IEEEpeerreviewmaketitle


%
%
%
%

%


\blfootnote{This work has been submitted to the IEEE for possible publication. Copyright may be transferred without notice, after which this version may no longer be accessible.}

\section{Introduction}
Since, in general, in computer networks multiple communications are simultaneously active,
the users have to share the limited network resources.
When sharing the network links, the users can either behave selfishly, meaning that each user strives to use as much resources as possible; alternatively, they can cooperate to increase the overall efficiency of the communication system. In communication networks, the \emph{price of anarchy} (which is the degradation of the system efficiency due to a selfish behavior) is rather high and the use of resource allocation protocols for an efficient sharing of the network resources is recommended.
Rate allocation problems in communication networks are commonly referred to as Network Utility Maximization (NUM) problems~\cite{kelly}. In this sort of problems a utility function is associated to each user in order to map the usage of the network resources to the obtained benefits. The goal of the communication system is then to share the network resources in such a way that the overall users' benefit is maximized.

\textcolor{black}{Most of the NUM optimization problems can be solved in a distributed way using primal and dual decomposition methods~\cite{tutorial}.
These decomposition methods form the basis of many network protocols used in communication systems~\cite{survey}. 
In NUM problems it is typically assumed that either the amount of available network resources is known, or that the users can measure some private congestion signals (e.g., packet losses or experienced delay) that can be used by the individual users to tune the transmitting rate and achieve the optimal rate allocation.
Although in real communication networks users have always access to a private congestion signal, it is not always true that these signals can be used by the individual users to adjust their transmitting rate to the optimal value. For instance, this problem arises clearly in the scenario of HTTP Adaptive Streaming (HAS)~\cite{stockhammer,decicco}. HAS represents nowadays the standard technology for video streaming over the internet and it employs as communication protocols HTTP over TCP. The key point is that HAS users sharing a common link tend to converge to a rate-fair allocation if they tune the rate according to their individual congestion measurements. As it is shown in~\cite{mine,turck}, such a rate-fair allocation is not efficient for video streaming and users should rather be coordinated at the application level in order to reach a higher quality of service.  The same problem may arise whenever a network application used by different users is forced to employ a specific congestion control algorithm (e.g., TCP), while it would be better to coordinate the transmitting rates to match a specific rate allocation.
If these network applications have no knowledge about the amount of available resources, the optimal rate allocation is hard to find, and classical NUM algorithms cannot help in this scenario. In this work, we aim at filling this gap and propose a way for extending the NUM framework to more general scenarios where resource constraints are not known a priori.}

Specifically we consider a set of users sharing different network links. The users want to transmit data across the network while optimizing the system efficiency. We assume that the system knows how the network nodes are connected but does not know the transmitting capacity of the links. We further assume that, when the overall transmitting rate exceeds the network capacity, only some of the network users may detect the congestion event. The goal is then to design a distributed overlay allocation method that tunes the sending rate of all the users, so that even the users that do not detect any congestion event can adjust their sending rate to their optimal value.

In more detail, we first separate the problem in two subproblems, one corresponding to a classical NUM problem and one corresponding to a resource inference problem. Due to the general and complex model assumed for the congestion signals the resource inference cannot be done efficiently using a passive method and the adoption of an active learning method is crucial in order to speed up the resources estimation. In the analyzed framework, performing active learning comes at the expenses of reducing the service provided to the users. 
In order to guarantee good system performance in both short and long term, we formulate the  problem as an \emph{optimal learning} problem~\cite{powell}. 
The original optimal learning problem is however computationally intractable, hence we carefully approximate the different mathematical steps and develop a decentralized algorithm to solve a simplified version of the problem. The algorithm distributes the operations among different processes, each associated to a single specific network link. 
These processes are not bounded to be executed on any specific network node, and only require communication among each other and with the different users, enabling a flexible deployment.
The experimental results show the effectiveness of the proposed method and the advantage that an optimal learning formulation offers with respect to a na\"ive strategy that greedily optimizes the immediate performance.

\textcolor{navy}{Motivated by similar arguments, some works, such as~\cite{gradienterror1,gradienterror2}, analyze how classical NUM algorithms are affected by potential noisy measurements of the feedback signals. In particular these studies consider the case where the measurements are corrupted by a biased noise which prevents the network from achieving the optimal rate allocation. The aim of our study compared to the latter ones is however different: whereas these works try to assess the optimality gap due to biased feedbacks, in our work we try to infer the available resources in order to implement an overlay NUM system that is then robust to biased feedback.}
We are not aware of any work on NUM that tries to build a user coordination system without having access to direct measures of the available resources. The only known work that attempts to solve a NUM problem with unknown constraints is~\cite{ribeiro}. The authors however assume to have access to measurements of the constraint vector perturbed by a zero mean noise. In our case we rather assume to never have direct access to the links capacities but only to indirect users congestion signal  measurements.
A somewhat similar problem is analyzed in~\cite{LCG}. In this NUM related work, the private congestion signals measured by the different users are heterogenous, leading to an inefficient rate allocation. The authors use a game theoretic framework in order to design a coordination system  based on local users' beliefs without any explicit user communication. However in this study it is assumed that the congestion signals that the user can measure are linearly coupled with the actions, i.e., the sending rates, of all the users. In our case instead, we do not make such assumption, and the congestion signal observed by some users could be completely insensitive to the actions of other users that employ the same network resources. Therefore we are able to cover a larger class of NUM problems.


This paper is organized as follows. In Section~\ref{sec:not_and_back} we provide some background related to NUM problems. In Section~\ref{sec:prob_form} we introduce the problem settings and state the problem formulation. In Section~\ref{sec:prop_sol} we describe how we approximate and solve our instance of the NUM problem. In Section~\ref{sec:full_alg} we summarize the proposed method. Results from computer simulations are provided in Section~\ref{sec:results}. Finally conclusions are provided in Section~\ref{sec:conc}.


\section{Background}\label{sec:not_and_back}
In this work we denote vectors with bold lowercase letters ($\v{x}$) and matrices with bold uppercase letters ($\v{X}$). In both cases the non-bold lowercase symbol with subscripts ($x_n$ or $x_{nm}$) denotes a single element of the vector (or matrix). The notation $(\v{x},y)$ denotes a tuple of different variables whereas $\{\v{x}_k\}$ denotes a collection of elements indexed by $k$.

\begin{figure}
\centering
\includegraphics[width=0.9\columnwidth]{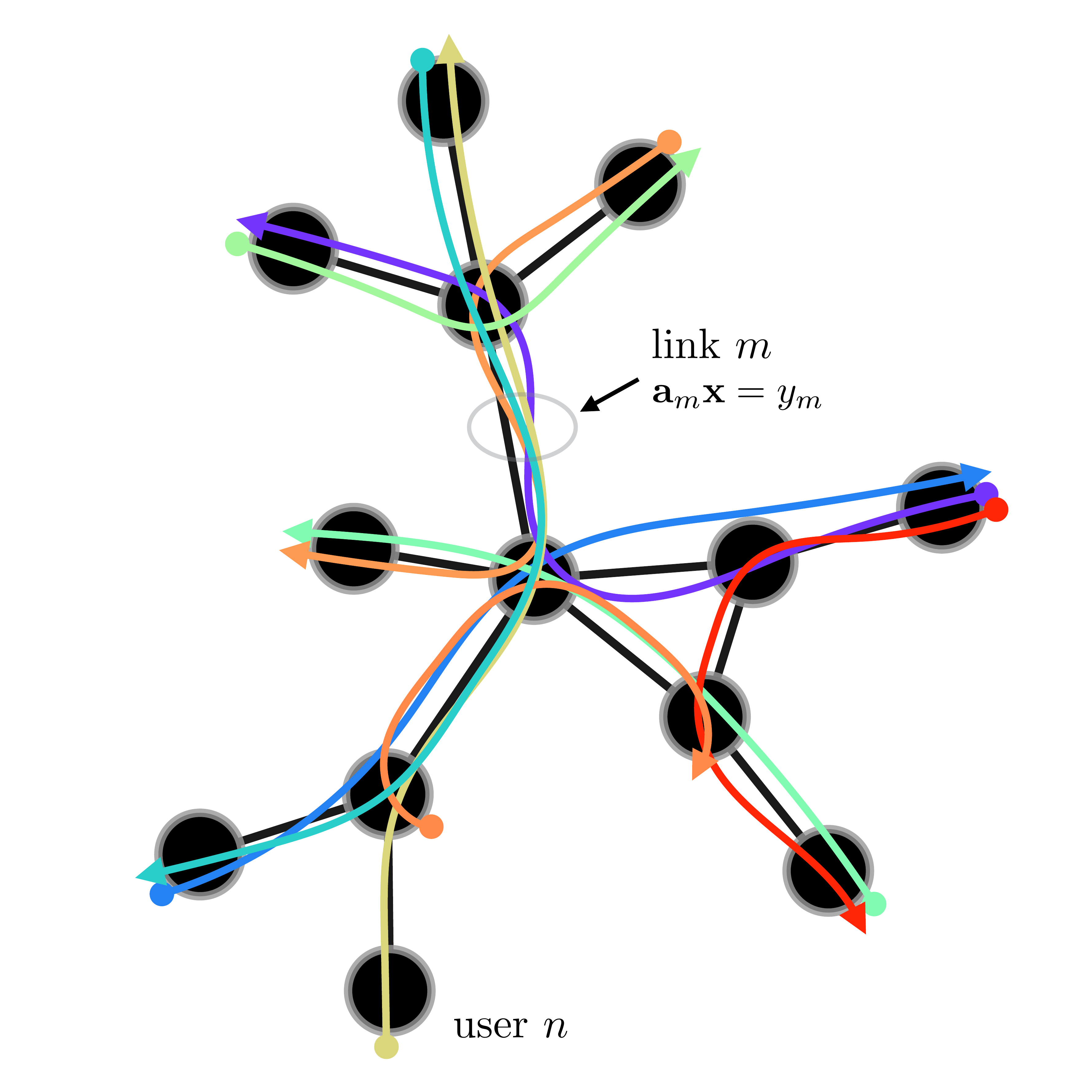}
\caption{Overview of the considered scenario. Each user resides on a node of the network and transmits data to another network node. The colored arrows represent the communication route of each user.}
\label{fig:network}
\end{figure}

As depicted in Fig.~\ref{fig:network}, we consider a set of $M$ network links shared by a set  of $N$ users. Each user transmits data between two nodes of the network. The subset of adjacent network links that connect the source node  and the destination node of a user communication forms the route taken by the user's data.
The routing information is embedded in the matrix $\v{A}$, which is a $M\times N$ binary matrix, where the element $a_{nm}$ is equal to $1$ if user $n$ employs link $m$, and $0$ otherwise.

The vector $\v{x} \in \mathbb{R}^N$ represents the transmitting rate of each user.
We associate to each user $n$ a utility function $u_n(x_n)$ that maps the transmitting rate of user $n$, i.e., $x_n$, to the resulting benefit. In NUM problems, utility functions are usually assumed to be strictly increasing smooth concave functions~\cite{low2002,fastnum}.
An efficient utilization of the resources is given by the solution to the following NUM problem:
\begin{equation}
\begin{aligned}
\underset{\v{x}}{\text{maximize}} \ \ &\ \mathcal{U}(\v{x})=\sum_{n=1}^{N} u_n(x_n)\\
\text{subject to} \ \ &\  \v{A}\v{x} \leq \v{b}\\
\ \ &\ \v{x} \in \mathcal{X},
\end{aligned}
\label{eq:NUM}
\end{equation}
where the vector $\v{b}$ represents the available resources, i.e., the capacities of the network links. $\mathcal{X}$ represents the domain of the utility functions, which usually coincides with the positive orthant, as negative transmitting rates are meaningless.
The optimization problem in Eq.~\eqref{eq:NUM} aims at maximizing the overall utility of the users subject to the limited availability of the resources. 
NUM problems are generally solved online with a distributed algorithm~\cite{survey,low2002}. 
The solving method basically consists in the design of a distributed closed loop control system whose equilibrium point is the optimal rate allocation of the NUM problem.
At each time step $t$, the users demand a certain amount of resources $\v{x}^t$ and observe a private feedback signal representing the constraint violation. The users then independently adapt  the request at step $t+1$ according to the private feedback received, converging to the equilibrium point.  

One possible way to solve the NUM problem in a distributed way relies on a dual decomposition~\cite{tutorial}.
The dual function of problem in Eq.~\eqref{eq:NUM} corresponds to:
\begin{equation}
g(\vg{\lambda},\v{b})=\underset{\v{x} \in \mathcal{X}}{\text{max}}\  \mathcal{U}(\v{x}) - \vg{\lambda}^\tran\left( \v{A}\v{x}-\v{b} \right) 
\label{eq:dual}
\end{equation}
where the vector $\vg{\lambda}$ represents the dual variables, or prices, associated to the $M$ constraints. The optimal point for the primal and dual variables $(\v{x}^\star,\vg{\lambda}^\star)$, can be found by iterating over the following steps:\begin{subequations}
\begin{empheq}[left={\empheqlbrace\,}]{align}
& \label{eq:primal_update} {x}^{t}_n = \underset{x' \in \mathcal{X}_n}{\text{arg max}} \{u_n(x') - x'\v{a}_n^\tran\vg{\lambda}^t \}\  & n=1...N\\ 
& \label{eq:dual_update} {\lambda_m^{t+1} } =  \Bigg( \lambda_m^t + \epsilon  \left( \v{a}_m\v{x}^{t} - b_m \right) \Bigg)^+\  & m=1...M,\end{empheq}\label{eq:solveNUM}\end{subequations}where $\epsilon$ is a parameter that controls the step size of the dual variables update and $()^+$ denotes the projection onto the positive orthant. The quantities $\v{a}_n^\tran\vg{\lambda}^t$ represent the private signals that are needed by each user in order to update the sending rate.
Eq.~\eqref{eq:solveNUM} represents a distributed feedback control loop whose equilibrium point corresponds to the optimal point of the NUM problem in Eq.~\eqref{eq:NUM}.
In the common settings of the NUM problem, it is usually assumed that either the link capacities $\v{b}$ are known a priori, so that Eq.~\ref{eq:dual_update} can be computed, or that the dual variables $\vg{\lambda^t}$ correspond to physical signals that can be measured by the individual users. For instance in delay-based congestion control algorithms the dynamics of the packet queuing delays match the dynamics of the dual variables, and the users can collect their private feedback signals by measuring the experienced delay.
How to achieve the optimal allocation efficiently, when the available resources are not known a priori and the private signals cannot directly be used to control the users sending rate, is an open problem that we aim to solve in this study.

\section{Problem Settings and Framework}\label{sec:prob_form}

\subsection{Problem Settings}\label{sec:prob_settings}
We want to solve the NUM problem of Eq.~\eqref{eq:NUM} when the exact value of the vector $\v{b}$ is unknown.
In order to find a solution to this problem we rely on the assumption that when the sending rates of the users exceed the available resources, the users can detect a congestion event. In the following we define more precisely what are the features of the congestion signals observed by the users, and define a probabilistic model that relates these signals to the link capacities. The model is then used by our rate allocation scheme to infer the constraint vector $\v{b}$.

We define the random binary variable $z^t_m$ to represent the occurrence of a congestion event on link $m$. Its probability is given by:
 \begin{equation}
 \begin{aligned}
p(z^t_m|b_m, y^t_m) =& \left[\sigma(\kappa(y^t_m - \rho b_m))\right]^{z^t_m}\cdot\\
				   & \left[1- \sigma(\kappa(y^t_m - \rho b_m))\right]^{1-z^t_m},
 \end{aligned}
\label{eq:z_def}
\end{equation}
where $\sigma()$ denotes the sigmoid function, $y^t_m$ corresponds to the sum of the users sending rates passing through  link $m$ at time $t$, $y^t_m=\v{a}_m^\tran \v{x}^t$. $\kappa$ and $\rho$ are positive scalar parameters that can be used to tune the steepness and the location of the sigmoid function.
 Eq.~\eqref{eq:z_def} tells us that the larger is the requested transmission rate for link $m$ the larger is the probability to face a congestion event.
By using the sigmoid function instead of a step function in Eq.~\eqref{eq:z_def} we can account for some possible noise in the network, e.g., transmission bursts and noisy estimates of $y_m$.
As the binary variable $z^t_m$ represents the occurrence of a congestion event on link $m$ at time $t$, the binary variable $v^t_n$ represents the detection of a congestion event on the route of user $n$ at time $t$ ($v^t_n=1$ when a congestion is detected by user $n$).

In this study we assume that the user congestion variables $\v{v}^t$ and link congestion variables $\v{z}^t$ are related by the following conditions:
\begin{itemize}
\item[\emph{a})] if $v^t_n=1$ then at least one variable $z^t_m$ with $a_{mn}=1$, has to be equal to one. This condition tells us that if all the links used by the user do not trigger any congestion, then the user cannot observe a congestion event.
\item[\emph{b})] If for all the users with  $a_{mn}=1$, the variables $v^t_n$ are zero,  then $z_m^t=0$. This condition tells us that if link $m$ triggers a congestion then one of the users employing link $m$ has to observe a congestion event.
\end{itemize}
If two vectors $\v{z}^t$ and $\v{v}^t$ do not verify the above conditions, then, they are not consistent and the probability to observe a pair of inconsistent vectors is zero.
\textcolor{navy}{
In order to model mathematically the above conditions we introduce the function $\Psi_{\v{v}^t}(\v{z}^t)$. This function is parametrized by a vector $\v{v}^t$, and maps a vector $\v{z}^t$ to the set $\{0,1\}$. Given a pair of vectors $\v{v}^t$ and $\v{z}^t$, $\Psi_{\v{v}^t}(\v{z}^t)$ is equal to one if the two vectors verify the above conditions and equal to zero otherwise.
More specifically the function is defined as follows:
\begin{equation}
\Psi_{\v{v}^t}(\v{z}^t) = \prod_{n:\ v_n^t=1} \psi_{n}^1 (\v{z}^t)  \prod_{m:\ \v{a}_m^\tran \v{v}^t=0}\psi_{m}^0 ({z}_m^t),\label{eq:full_z_const}
\end{equation}
with:
\begin{equation}
\psi_{n}^1 (\v{z}^t) = 1-  \prod_m (1- a_{nm}z^t_m) , \ \ \psi_{m}^0 ({z}_m^t) = 1-  z^t_m.
\label{eq:z_const_2}
\end{equation}
The factors $\psi_{m}^0$ correspond to the above condition \emph{b}) whereas the factors $\psi_{n}^1$ are associated to condition $\emph{a}$).
Note that the vector $\v{v}^t$ selects the factors $\psi_{n}^1$ and $\psi_{m}^0$ that are active in full term $\Psi_{\v{v}^t}$.}

\begin{figure}
\centering
\includegraphics[width=1.\columnwidth]{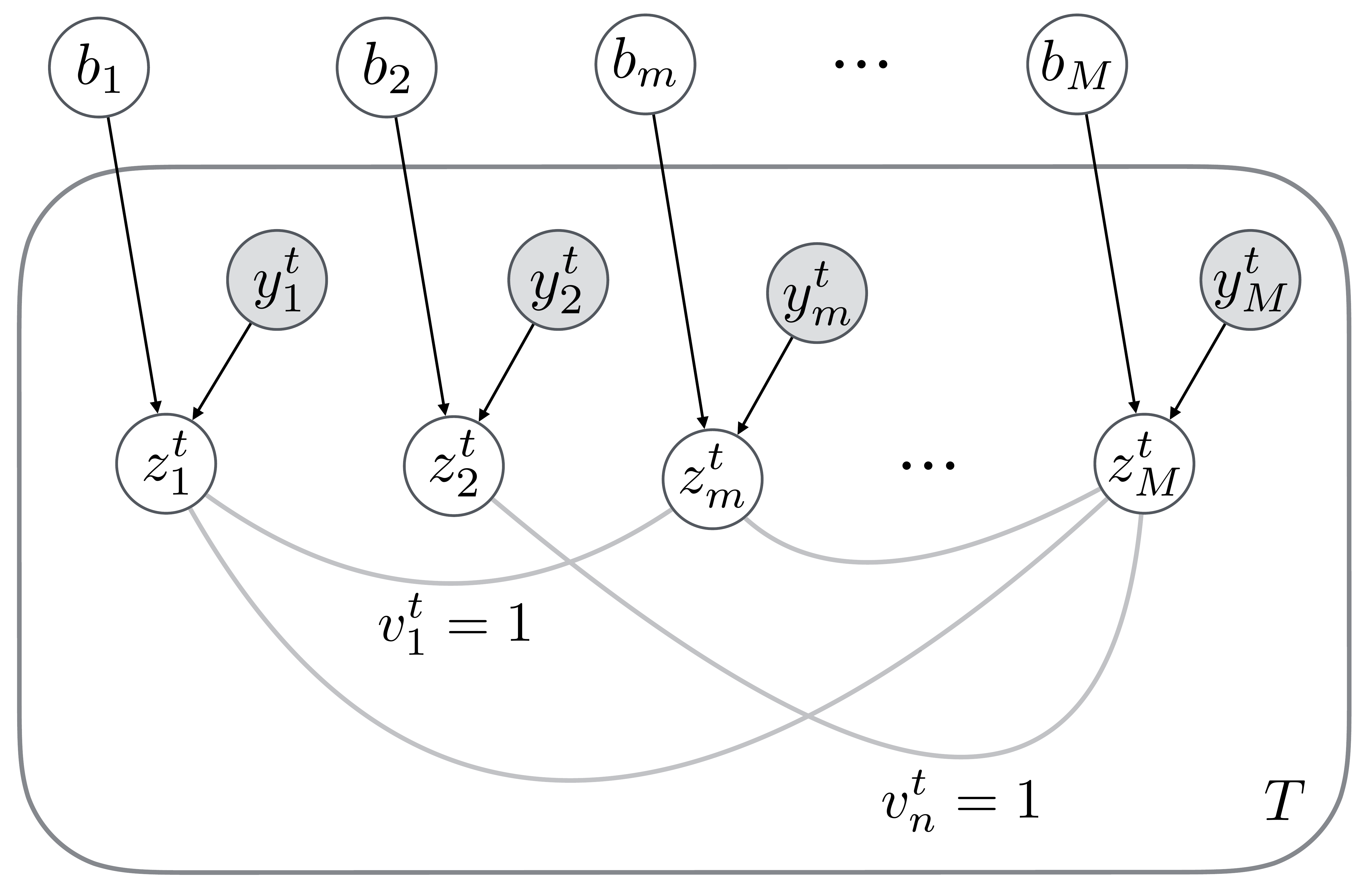}
\caption{Graphical model of the considered scenario. Grey variable nodes represent quantities that are observed at each step $t$.}
\label{fig:graphmodel}
\end{figure}

What we actually observe at each iteration $t$ is not the link congestion vector $\v{z}^t$, which represents a latent variable in our model, but the user congestion signals $\v{v}^t$ and the users rates $\v{x}^t$ (since we assume to know the routing matrix the vector $\v{y}^t$ is also known). As shorthand we denoted by $\mathcal{D}^t$ the observed data at time $t$: $\mathcal{D}^t = (\v{y}^t,\v{v}^t)$.
What we aim to do is to use the observed variables in order to infer the value of the constraint vector $\v{b}$, which is the quantity we are interested in, in order to find the optimal rate allocation.
We can represent the above probabilistic model using the graphical model of Fig.~\ref{fig:graphmodel}. As can be seen the $\v{z}^t$ variables are generated from the vectors $\v{b}$ and $\v{y}^t$ but they are related to each other depending on the value of the observed vector $\v{v}^t$.
Combining Eq.~\eqref{eq:z_def}-\eqref{eq:z_const_2} for all the observation $t$ we obtain:
\begin{equation}
p(\{\v{z}^t\}|\v{b},\{\mathcal{D}^t\}) =\frac{1}{Z} \prod_{t}  \Psi_{\v{v}^t}(\v{z}^t) p(\v{z}^{t}|\v{b},\v{y}^{t}),
%
\label{eq:zed_full}
\end{equation}
where $p(\v{z}^{t}|\v{b},\v{y}^{t})=\prod_m p({z}_m^{t}|{b}_m,{y}_m^{t})$ and $Z$ corresponds to a normalization constant required to obtain a valid posterior distribution.
Denoting with $p^0(\v{b})$ the prior knowledge on the parameters $\v{b}$ we can write down the joint distribution between the latent variables $\{\v{z}^t\}$ and the parameters $\v{b}$:
\begin{equation}
p(\{\v{z}^t\},\v{b}|\{\mathcal{D}^t\}) = p(\{\v{z}^t\}|\v{b},\{\mathcal{D}^t\})p^0(\v{b}).
%
\label{eq:joint_full}
\end{equation}
Finally the posterior on $\v{b}$  can be obtained by marginalizing out all the latent variables $\{\v{z}^t\}$, leading to:
\begin{equation}
p(\v{b}|\{\mathcal{D}^t\}) = \sum_{\{\v{z}^t\}} p(\{\v{z}^t\},\v{b}|\{\mathcal{D}^t\}).
%
\label{eq:post_b}
\end{equation}
The factor graph associated to Eq.~\eqref{eq:zed_full}-\eqref{eq:joint_full} is depicted in Fig.~\ref{fig:factor_graph}.

In our scenario we sequentially collect  the observable data $\mathcal{D}^t$, therefore we can see the belief on $p(\v{b})$ as something evolving with $t$:
\begin{equation}
p^{t-1}(\v{b},\v{z}^t|\mathcal{D}^{t}) = \frac{1}{Z^t} \Psi_{\v{v}^t}(\v{z}^t) p(\v{z}^{t}|\v{b},\v{y}^{t}) p^{t-1}(\v{b}),
\label{eq:joint_seq} 
\end{equation}

\begin{equation}
p^{t}(\v{b}) = p^{t-1}(\v{b}|\mathcal{D}^{t}) = \sum_{\v{z}^{t}} p^{t-1}(\v{b},\v{z}^{t}|\mathcal{D}^{t}).
\label{eq:b_marginal_seq}
\end{equation}
We can think of $p^{t}(\v{b})$ as the prior at time $t$, which is equal to the posterior at time $t-1$ ($p^{t-1}(\v{b}|\mathcal{D}^{t})$).

In this subsection we have formalized the user congestion signal $\v{v}$ along with a probabilistic model able to generate a posterior distribution on the constraint vector $\v{b}$ using the observed quantities $\v{v}$ and $\v{y}$.

\begin{figure}
	\centering
	\includegraphics[width=1.\columnwidth]{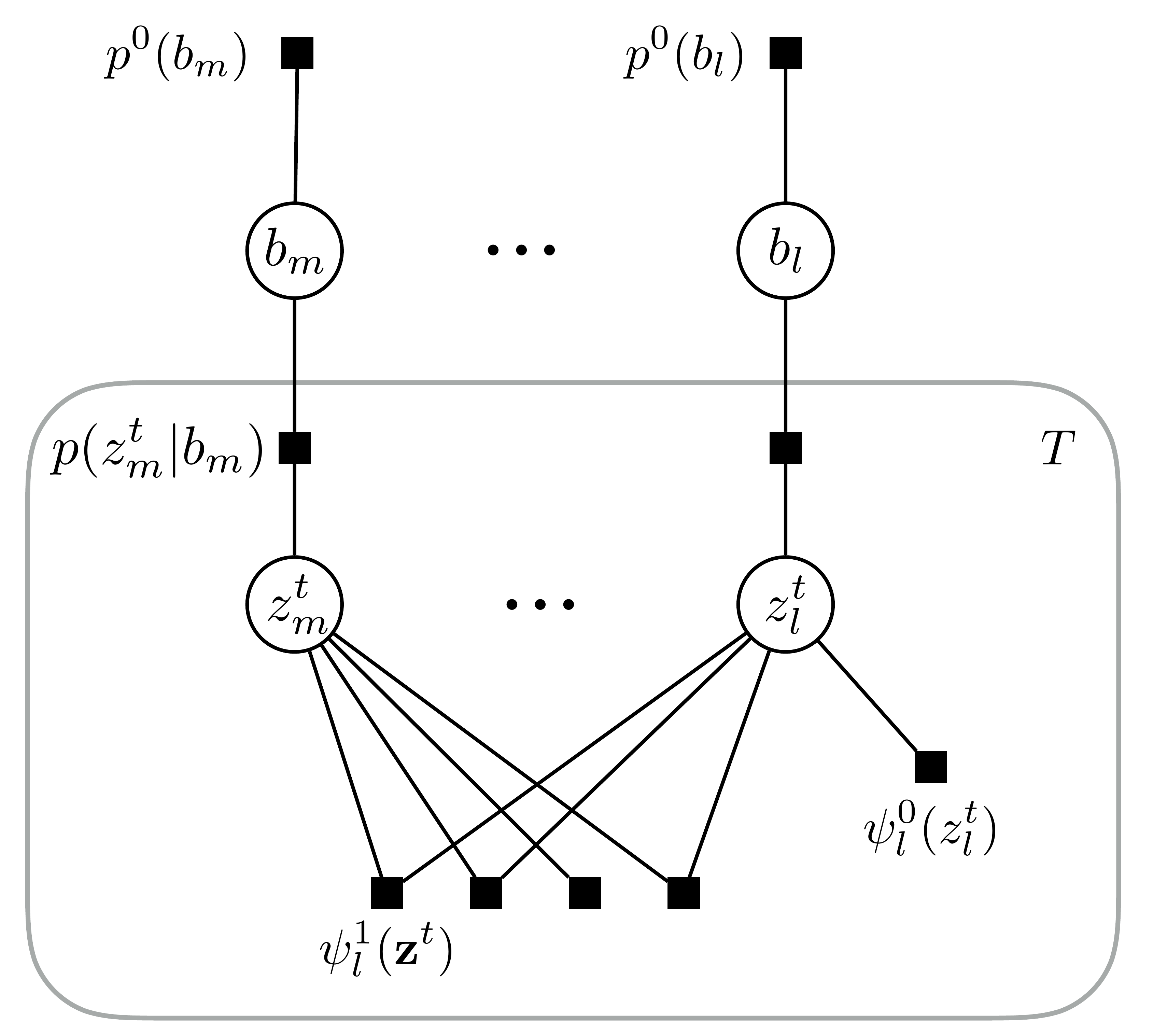}
	\caption{Factor graph of one single observation $t$ of the graphical model depicted in Fig.~\ref{fig:graphmodel}. Each factor $\psi_{n}^1 (\v{z}^t)$ forces at least one of the connected variables $\v{z}^t$ to be one. The factor $\psi_{n}^0 ({z}_l^t)$ force $z_l^t$ to be zero.}
	\label{fig:factor_graph}
\end{figure}

\subsection{System Architecture}
Analogously to the classical NUM framework in Section~\ref{sec:not_and_back}, we need to design a closed loop algorithm that selects at each step $t$ a certain rate vector $\v{x}^t$. It then observes the congestion feedback signal $\v{v}^t$, and uses it to compute a new value of the sending rate $\v{x}^{t+1}$. This process should continue till convergence, ideally to the optimal allocation $\v{x}^\star$ of Problem~\eqref{eq:NUM}.

\textcolor{navy}{There is no unique solution to the design of such feedback control system. One possible option consists in having a single loop algorithm where the allocation method at each iteration $t$ computes under a defined policy a new allocation vector $\v{x}^{t+1}$.
While this design choice ideally allows to achieve better performance, as it does not impose any particular structure on the control system, it also poses several design challenges. The controller should, in fact, find a map between the prior belief $p(\v{b})$, the users' requests $\{\v{x}^t\}$ and the feedbacks $\{\v{v}^t\}$ (along with the probabilistic model that relates $\mathcal{D}^t$ to $\v{b}$), to find the rate allocation $\v{x}^{t+1}$. Due to the large dimension of the input and output space this design choice is unpractical}

\begin{figure}
\centering
\includegraphics[width=1.\columnwidth]{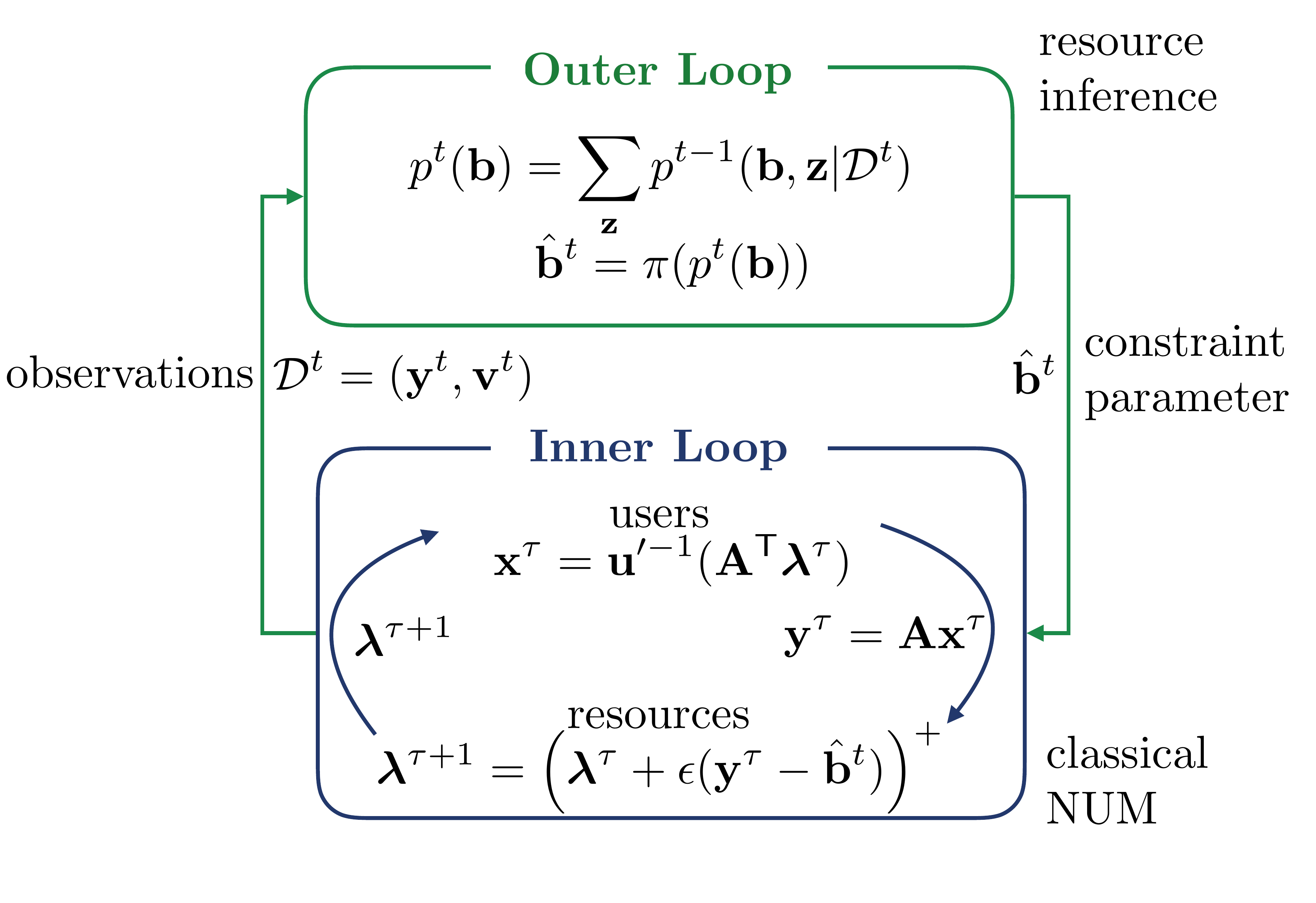}
\caption{High-level block diagram of our system. The method is composed by two separate loops. The inner loop represents a simple primal-dual distributed algorithm for solving the NUM problem (e.g., Eq.~\eqref{eq:solveNUM}). The outer loop is the  method proposed in this work, see Section~\ref{sec:prop_sol}, for the selection of the sequence of constraints $\{\hat{\v{b}}^t\}$ to be used as input parameter for the inner loop.}
\label{fig:system_overview}
\end{figure}

\textcolor{navy}{An alternative design choice consists in an adaptive control architecture~\cite{wittenbook}, where we separate the resource allocation method in two separate loops. An inner loop responsible for the users coordination, where $\v{b}$ is seen as a controller parameter, and an outer loop responsible for the inference of the $\v{b}$ parameter. A high-level block diagram of this system is depicted in Fig.~\ref{fig:system_overview}. In this case the design of the controller is much simpler as the user rate allocation process is separated from the resource inference.
The inner loop simply has to solve at step $t$ an instance of the NUM problem with the available resources set by the vector $\hat{\v{b}}^{t}$. In this case any method capable to solve the NUM problem, as for instance the one in Eq.~\eqref{eq:solveNUM}, can be used.
The outer loop is responsible for the inference of the true $\v{b}$ vector. It receives as input the samples $\mathcal{D}^t$, and, based on the current belief $p^t(\v{b})$, it then selects the best value of $\hat{\v{b}}^{t}$ under some defined policy. Note that the belief $p^t(\v{b})$ represents a hyperstate of the controller that evolves according to Eq.~\eqref{eq:joint_seq}-\eqref{eq:b_marginal_seq}. Differently from the inner loop there is no off-the-shelf solution for the outer loop controller, the design of such subsystem is the main focus of this work.
As final remark, in our design we consider to update the outer loop, and compute a new parameter $\hat{\v{b}}$, after the inner loop converges to equilibrium. In this case, the observed data $\mathcal{D}^{t+1}$ corresponds to the equilibrium point of the inner loop at iteration $t$, which is the optimal point of the NUM problem of Eq.~\eqref{eq:NUM} when the $\v{b}=\hat{\v{b}}^{t}$, i.e., $\v{y}^t=\v{y}^\star(\hat{\v{b}}^{t-1})$.}


\subsection{Problem Formulation}\label{subsec:probform}
In this subsection we define the optimization problem that we ideally aim to solve in order to design the outer loop controller.

\textcolor{black}{The design of the outer loop controller corresponds to finding a policy $\pi(\cdot)$ that maps the current belief on the constraint vector $\v{b}$ into a value of the constraint vector to be used as parameter for the inner loop ($\pi : p(\v{b}) \rightarrow \hat{\v{b}}$) while maximizing the performance of the system. 
We can measure the performance of the system at each step $t$ as the expected optimality gap when using as constraint vector for the classical NUM algorithm of the inner loop $\hat{\v{b}}^t$ instead of the true unknown value $\v{b}$. In order to do this, we introduce a loss function $L(\v{b},\hat{\v{b}})$. Ideally the loss function models the difference in overall users' utility attained by using the estimate $\hat{\v{b}}$ instead of the true value $\v{b}$. 
Unfortunately, an exact evaluation of the optimality gap is highly expensive in terms of computations since it would require to solve the NUM problem of Eq.~\eqref{eq:NUM} twice. Instead we can consider a simpler and rather classic loss function such as the squared $l^2$ distance:
\begin{equation}
L(\v{b},\hat{\v{b}}) = || \v{b} - \hat{\v{b}} ||_2^2.
\label{eq:loss_used}
\end{equation}
The idea is that the closer $\hat{\v{b}}$ is to the true vector $\v{b}$ the smaller the optimality gap.
Using the belief $p(\v{b})$ and the loss function we can then define the risk as the expected value of the loss function:
\begin{equation}
\mathcal{R}(\hat{\v{b}},p(\v{b})) = \mathbb{E}_{p(\v{b})} \left[ L(\v{b},\hat{\v{b}}) \right].
\label{eq:risk}
\end{equation}
The risk corresponds to our performance metric at each step $t$ and it represents the expected suboptimality of the rate allocation algorithm when using a resource estimate equal to $\hat{\v{b}}$.}
Having defined the risk for a single step, we can easily extend the same metric to an entire sequence of belief-estimates pairs $\{(p^t(\v{b}),\hat{\v{b}}^t)\}$:
\begin{equation}
\mathcal{R}^\infty(\{(\hat{\v{b}}^t,p^t(\v{b}))\}) = \sum_{t=0}^{\infty} \gamma^t  \mathcal{R}(\hat{\v{b}}^t,p^t(\v{b})),
\label{eq:cost_to_go}
\end{equation}
where $\gamma \in [0,1)$ represents a discount factor that permits to compute the cumulative risk over an infinite time horizon. 

Maximizing the expected long term performance of our system is equivalent to minimizing the long term risk of being suboptimal, corresponding to Eq.~\eqref{eq:cost_to_go}. Note that consecutive beliefs  $p^t(\v{b}))$ are dependent;  more precisely, they evolve according to Eq.~\eqref{eq:joint_seq}-\eqref{eq:b_marginal_seq}, and the evolution depends on the value of the observed data $\mathcal{D}^t$.
By combining the different elements we can formulate an optimal learning problem for the long term risk minimization:
\begin{equation}
\begin{aligned}
\underset{\pi(\cdot)}{\text{minimize}} \ \ &\ \mathbb{E}_{p\left(\{\mathcal{D}^{t}\}|\{\pi(p^{t-1}(\v{b}))\}\right)} \left[ \sum_{t=0}^{\infty} \gamma^t  \mathcal{R}(\pi(p^t(\v{b})),p^t(\v{b})) \right]\\
\text{subject to} \ \ &\ p^{t}(\v{b}) = \sum_{\v{z}^{t}} p^{t-1}(\v{b},\v{z}^{t}|\mathcal{D}^{t}),
\end{aligned}
\label{eq:fulloptimal}
\end{equation}
where $p\left(\{\mathcal{D}^{t}\}|\{\pi(p^{t-1}(\v{b}))\}\right)$ corresponds to the probability of observing a sequence $\{\mathcal{D}^{t}\}$ for $t=1...\infty$ given that a constraint vector $\hat{\v{b}}^t=\pi(p^{t-1}(\v{b}))$ is used at step $t$ for the inner control loop, marginalized over the initial belief on $\v{b}$, i.e.,
\begin{equation}
p(\{\mathcal{D}^{t}\}|\{\pi(p^{t-1}(\v{b}))\})=\prod_t \int p(\mathcal{D}^{t}|\pi(p^{t-1}(\v{b})),\v{b})p^{t-1}(\v{b}) d\v{b}.
\label{eq:predict_D}
\end{equation}
Since the observed data $\mathcal{D}^{t}$ is a random variable, the transition from $p^{t-1}(\v{b})$ to $p^{t}(\v{b})$ is also stochastic. As a result we need to average the objective function of Eq.~\eqref{eq:fulloptimal} over the possible transitions.

The discount parameter $\gamma$ is a free parameter that represents how much we care about the future performance. 
The choice $\gamma=0$ represents a greedy strategy where at each step the controller would select $\hat{\v{b}}^t$  in order to minimize the immediate risk without caring about the future steps.
Taking into account the long term risks basically is what allows to perform active learning. Since different data points $\mathcal{D}^{t+1}$ lead to different future beliefs on $\v{b}$ (and beliefs with a smaller variance are beneficial because they result in a smaller risk), it is important to select $\hat{\v{b}}^t$ that reduces both, the current risk and the uncertainty of the future beliefs. As we will see, in our problem values of $\hat{\v{b}}^t$ that reduce the current risk, do not coincide with those that reduce the expected value of the future risk. Using an optimal learning formulation we are able to explicitly take into account this tradeoff and perform a smart choice for  $\hat{\v{b}}^t$ at each step $t$.



\section{Approximate Solution}\label{sec:prop_sol}
\textcolor{navy}{In this section we describe how it is possible to approximate the problem in Eq.~\eqref{eq:fulloptimal} in order to be able to find an effective solution. We first limit the time horizon over which we optimize, we then describe how it is possible to approximate the true posterior on $\v{b}$ using deterministic approximate inference, and finally we introduce a mean field approximation  in order to deal with large network systems.}

\subsection{Receding Horizon Adaptation}
In order to find the optimal policy for the closed loop problem of Eq.~\eqref{eq:fulloptimal} we typically need to compute the state value function of our system using a dynamic programming approach. However, due to the infinite dimension of the state space and the required expectation operations, this method quickly runs into computational problems.
One way to work around the complexity problem consists in approximating the closed loop problem of Eq.~\eqref{eq:fulloptimal} with a sequence of receding horizon open loop problems. As also suggested in~\cite{powell}, a rough but effective approximation for optimal learning problems corresponds to computing at each time step $t$ the estimate $\hat{\v{b}}^t$ that minimizes the long term risk as step $t$ would be the last step we were allowed to learn and modify our belief. Then, after taking decision  $\hat{\v{b}}^t$ and observing $\mathcal{D}^{t+1}$, the belief is updated and the same problem is solved again with the new belief. 

Using this approach at each step $t$ we aim at finding the estimate $\hat{\v{b}}^t$ given the current belief $p^t(\v{b})$ that minimizes the risk of the current step $t$ plus an additional term corresponding to the discounted infinite sum of the expected minimum immediate risk at step $t+1$. Hence, we aim at solving the following optimization problem:
\begin{equation}
\begin{aligned}
\underset{\hat{\v{b}}'}{\text{minimize}} \ \ &\ \mathcal{R}(\hat{\v{b}}',p^t(\v{b})) +\frac{\gamma}{1-\gamma}\mathbb{E}_{p^{t}(\mathcal{D}|\hat{\v{b}}')} \left[ \mathcal{R}^\star(p^{t}(\v{b}|\mathcal{D})) \right]
\end{aligned}
\label{eq:MPC_like}
\end{equation}
where $\mathcal{R}^\star(p^{t}(\v{b}|\mathcal{D}))$ corresponds to the minimum risk for a belief equal to the posterior distribution.
The receding horizon approach, of Eq.~\eqref{eq:MPC_like}, has largely simplified the problem to solve, compared to Eq.~\eqref{eq:MPC_like}, mainly for two reasons: \emph{i}) being open loop we optimize over a vector of dimension $M$ rather than the infinite dimension policy $\pi$, and \emph{ii}) by limiting the learning horizon to one step we only need to compute the expectation over a single observation $\mathcal{D}$.

In the next subsection we describe how to use approximate inference methods in order to find a convenient expression of the belief $p^t(\v{b})$, which is necessary for the computation of the risk. We then deepen into the details of an approximate solution of the problem in Eq.~\eqref{eq:MPC_like} for large network systems.

\subsection{Approximate Posterior Inference}\label{approx_inference}
The probability distribution $p^t(\v{b})$ represents the belief on the parameter $\v{b}$ given our original prior $p^0(\v{b})$ and all the observations collected so far $\{\mathcal{D}^t\}$. The distribution $p^t(\v{b})$ is required to evaluate the risk of different estimates $\mathcal{R}(\hat{\v{b}},p^t(\v{b}))$. This operation involves averaging over all the possible values of $\v{b}$ and it is in general computationally expensive.
There are mainly two possible ways to compute expectations under a fixed distribution: deterministic approximate inference methods and sampling methods. In this work we  use the deterministic approximate methods as they are considered faster and require less computations to obtain an approximate result~\cite{minka_orig}. These methods attempt to minimize a divergence measure between the true posterior, $p(\v{b}|\{\mathcal{D}^t\})$, and a second distribution, $q(\v{b})$, which belongs to a fixed and predefined distribution family. The key point is that computing expectation is much simpler over the target distribution than over the original one and can often be done analytically. More specifically we implement the Expectation Propagation (EP) algorithm~\cite{minka_orig} on the graphical model of Fig.~\ref{fig:graphmodel}, using as target distribution for the parameter vector $\v{b}$ a fully factorized distribution $q(\v{b})$ composed of $M$ univariate lognormal distributions. The reasons for using this specific distribution is twofold: \emph{i}) the lognormal distribution has positive support, like the possible values of the link capacities, \emph{ii}) as required by the EP algorithm  it belongs to the exponential family. Note that considering a multivariate lognormal distribution with correlated components would certainly lead to a more accurate approximation of the true distribution, but it also has a higher storage and computational cost (the storage cost of the second order moments of the $M$ univariate distributions grows linearly with $M$, whereas, for the multivariate distribution it grows quadratically with $M$).


The EP algorithm with a fully factorized target distribution is similar to the \emph{loopy belief propagation}~\cite{LBP} algorithm and basically corresponds to an iterative distributed algorithm where information is exchanged among adjacent factor nodes and variable nodes on the factor graph.
\textcolor{navy}{The common intuition behind belief propagation algorithms is the following. Each factor node represents a bond among the $\mathcal{N}$ neighbor variable nodes in the form of a function of the $\mathcal{N}$ variables. For a given factor node, when the value of the $\mathcal{N}-1$ adjacent variable nodes is set, the factor node function can be used to produce an \emph{opinion} (belief) on the value of the left out variable node. The overall belief on a variable node can then be obtained by combining all the beliefs from all its adjacent factor nodes.}
A more technical and thorough description of the belief propagation algorithm and the EP algorithm goes beyond the scope of this work,
we refer the interested reader to the following works for further reading~\cite{minka_orig,bishop,MPDM}. Moreover, for a specific description of the implementation on the factor graph of Fig.~\ref{fig:factor_graph}, we refer the reader to Appendix~\ref{appA}.

As we consider a fully factorized target distribution, we basically associate to each variable node of the factor graph, depicted in Fig.~\ref{fig:factor_graph}, a univariate distribution with tunable parameters. For the $z$ nodes, this distribution is simply a Bernoulli distribution; whereas for the $b$ nodes as mentioned earlier, the associated distribution is the lognormal distribution. The iterative exchange of information among the nodes changes the parameters of these distributions till they converge to the value that minimizes a divergence measure with respect to the true distribution $p^t(\v{b})$.
After convergence, the outcome for the $\v{b}$ parameter is a set of $M$ univariate lognormal distributions that resemble the true belief $p^t(\v{b})$. At this point instead of computing the risk using the highly complex distribution $p^t(\v{b})$ we can use the approximate distribution $q(\v{b})$, which, being of a simple form, allows for a closed form expression of the risk $\mathcal{R}(\hat{\v{b}},q(\v{b}))$.

\subsection{Mean Field Approximate Solution}\label{optimize_future}
We now focus on the solution of problem of Eq.~\eqref{eq:MPC_like}.
Considering the quadratic loss introduced in Subsection~\ref{subsec:probform} and the factorized approximate belief $q(\v{b})$, we can express the risk $\mathcal{R}(\hat{\v{b}},q(\v{b}))$ in a simple form:
\begin{equation}
\begin{aligned}
\mathcal{R}(\hat{\v{b}},q(\v{b})) &= 
 \int ||\hat{\v{b}}-\v{b}||_2^2 q(\v{b}) d\v{b}\\
&= \sum_{m=1}^M \int  (\hat{b}_m-b_m)^2 q(b_m) d{b_m}\\
&= \sum_{m=1}^M \left( (\hat{b}_m-\mu_m)^2+\sigma_m^2 \right), 
\label{eq:risk_final}
\end{aligned}
\end{equation}
where $(\mu_m,\sigma_m^2)$ represents the mean and the variance of the distribution $q(b_m)$. The last expression of Eq.~\eqref{eq:risk_final} consists of a sum of $M$ parts, each corresponding to a single network link. Each part is composed by a sum of two terms: the square of the deviation from the mean of $q(b_m)$ plus the variance of $q(b_m)$. 
If we substitute Eq.~\eqref{eq:risk_final} in Eq.~\eqref{eq:MPC_like} we obtain an objective function composed by two parts, the squared $l^2$ distance of $\hat{\v{b}}$ from the mean of the current belief (immediate risk) plus a term that depends on the future belief (future risk).  Considering a posterior distribution computed by applying the EP algorithm and using the current $q(\v{b})$ as the true prior distribution, the second part of the objective function in  Eq.~\eqref{eq:MPC_like} corresponds the sum of the  variances of the future belief:
\begin{equation}
\begin{aligned}
\mathbb{E}_{p(\mathcal{D}|\hat{\v{b}})} \left[ \mathcal{R}^\star(p(\v{b}|\mathcal{D})) \right] &\simeq
\mathbb{E}_{p(\mathcal{D}|\hat{\v{b}})} \left[ \mathcal{R}^\star(q(\v{b})) \right] \\ & = \mathbb{E}_{p(\mathcal{D}|\hat{\v{b}})} \left[ \sum_m{\sigma^{2\prime}_m} \right],
\label{eq:exp_post_var}
\end{aligned}
\end{equation}
where $\sigma^{2\prime}_m$ denotes the posterior variance for parameter $b_m$, and we leveraged the fact that the risk is minimized when $\hat{b}_m=\mu_m$.
The computation of the future posterior variance after observing data $\mathcal{D}$ can be done by simply running loopy belief propagation, similarly to Subsection~\ref{approx_inference}. However, the difficulties reside in taking the expectation over all the possible observations $\mathcal{D}$. This operation is complicated for two reasons: \emph{i}) the number of possible combinations of the vector $\v{v}$ grows exponentially with the number of users $N$, \emph{ii}) we do not actually assume to have a generative model for the observations $\v{v}$.
As a workaround for these two impediments, we propose to adopt a mean field approximation of the network, and to consider a worst case scenario for the observations $\v{v}$.
The main motivation behind this approach is the following. Finding the true optimal constraint vector $\hat{\v{b}}$ is extremely complicated. Even for the approximate receding horizon problem of Eq.~\eqref{eq:MPC_like}, it would involve a large amount of computations, which are difficult to handle for large $M$ and $N$.
Therefore we ask whether we can optimize each entry of $\hat{\v{b}}^t$  independently by considering a mean interaction of all the network links. 

\textcolor{black}{As mean field approximation we consider a single network link with a lognormal distribution $q(\tilde{b})$ with mean and variance equal to $(\tilde{\mu},\tilde{\sigma}^2)$. We can write down the mean field version of Eq.~\eqref{eq:MPC_like} as:
\begin{equation}
\underset{\hat{{b}}}{\text{minimize}} \ \ (\hat{{b}}- \tilde{\mu})^2 +\frac{\gamma}{1-\gamma}\mathbb{E}_{p(\tilde{\mathcal{D}}|\hat{{b}})} \left[ \tilde{\sigma}^{2\prime} \right].
\label{eq:MPC_like_MF}
\end{equation}
The above equation simply represents the optimal learning formulation for a single mean field link. The quantity $\tilde{\mathcal{D}}=(\tilde{y},\tilde{\v{v}})$ represents the observed data related to the mean field link.
We now need to define an approximate relation between the parameter $\hat{{b}}$ and the future link rate $\tilde{y}$. The constraint vector $\hat{{b}}$ is used by the inner loop algorithm responsible for solving the classical NUM problem. Because of the shape of the utility functions, the users strive to utilize the resources as much as possible tending to make the constraints tight. As a result, we can consider, as a first approximation, $\tilde{y}=\hat{{b}}$.}

\textcolor{black}{
We now need to find an approximate expression for how the posterior variance of the mean field link $\tilde{\sigma}^{2\prime}$ varies with respect to $\mathcal{D}$. 
In order to proceed we need to consider separately the cases where the mean field link triggers a congestion, $\tilde{z}=1$, and  when it does not,  $\tilde{z}=0$.}

\subsubsection*{case $\tilde{z}=1$}
When the mean field link triggers a congestion event we know for sure that there must be at least one observation $\tilde{v}=1$ (see Subsection~\ref{sec:prob_settings}). This observation corresponds to a factor node in the factor graph that connects the mean field link to other network links, see Fig.~\ref{fig:mean_field}. If the average route length of the network is ${L}_R$, then the factor node is connected on average to the mean field link plus other  ${L}_R-1$ links. In this case the posterior variance of $\tilde{b}$, denoted by ${\tilde{\sigma}}^{2\prime}$, corresponds to the variance of the following distribution:
\begin{equation}
\begin{aligned}
{p'}(\tilde{b}) =& \bigg(\Big(1-\prod_{l < L_R-1}p(z_l=0|y_l) \Big)p(\tilde{z}=0|\tilde{y},\tilde{b})\\ 
& + p(\tilde{z}=1|\tilde{y},\tilde{b})\bigg)q(\tilde{b})/Z_{\text{norm}},
\label{eq:post_var_1}
\end{aligned}
\end{equation}
where $Z_{\text{norm}}$ is a normalizing constant, and 
\begin{equation}
p(z_l=0|y_l) =\int d{b} p({z}_l=0|{b_l},{y}_l) p({b}_l).
\end{equation}
The factor 
\begin{equation}
1-\prod_{l < L_R-1}p(z_l=0|y_l),
\label{eq:factor_post_var_1}
\end{equation}
of Eq.~\eqref{eq:post_var_1} is equal to the probability to trigger a congestion event by the other ${L}_R-1$ links composing the route. If the probabilities $p(z_l=0|y_l)$ are small $\forall l$ then it is easy to see that the variance of $p'(\tilde{b})$ is basically equal to the current one $\tilde{\sigma}^{2\prime} = \tilde{\sigma}^2$. The only way to achieve a gain in the posterior variance is to have a high probability $p(z_l=0|y_l)$ for all the other links composing the route. However, note that in the mean field model all the links are supposed to adopt the same strategy, as a result if $p(z_l=1|y_l)\simeq 0$ for all the links, it is also true that $p(\tilde{z}=1|\tilde{y})\simeq 0$ for the mean field link. In this case the considered event of a congestion by the mean field link would be extremely unlikely to happen, and by extension also the variance reduction.
One way to bypass the problem is to consider that the network links can belong to two different classes: class A and B. Links of class A have a probability of triggering a congestion event equal to $p_{\text{A}1}$, whereas links of class B have probability $p_{\text{B}1}$; more specifically $p_{\text{A}1}$ (similarly for $p_{\text{B}1}$) is defined as
\begin{equation}
p_{\text{A}1} = p(z=1|{y} = \hat{{b}}_\text{A})= \int d{b} p({z}|{b},{y}) p({b}) = 1 - p_{\text{A}0}.
\label{eq:p_A1}
\end{equation}
If we consider $p_{\text{A}1} > p_{\text{B}1} \simeq 0$, with the mean field link belonging to class A, and the other links composing the route belonging to class B, then Eq.~\eqref{eq:post_var_1} can actually lead to drastically change the mean field posterior belief.
By using the class notation Eq.~\eqref{eq:post_var_1} becomes
\begin{equation}
\begin{aligned}
p'(\tilde{b}) =& \Big(\left(1-p_{\text{B}0}^{{L}_R-1}\right)p(\tilde{z}=0|\tilde{y}_A,\tilde{b})\\ 
& + p(\tilde{z}=1|\tilde{y}_A,\tilde{b})\Big)q(\tilde{b})/Z_{\text{norm}},
\label{eq:post_var_1_valid}
\end{aligned}
\end{equation}
In the worst case scenario of a single user feedback $v$ for the mean field link, the probability of having such route is equal to $(1-\alpha)^{{L}_R-1}$. In all the other cases, i.e., when there is another link of the route that belongs to class A, we consider to have a posterior distribution with the same variance of the current belief.

\subsubsection*{case $\tilde{z}=0$}
\begin{figure}
\centering
\includegraphics[width=1.\columnwidth]{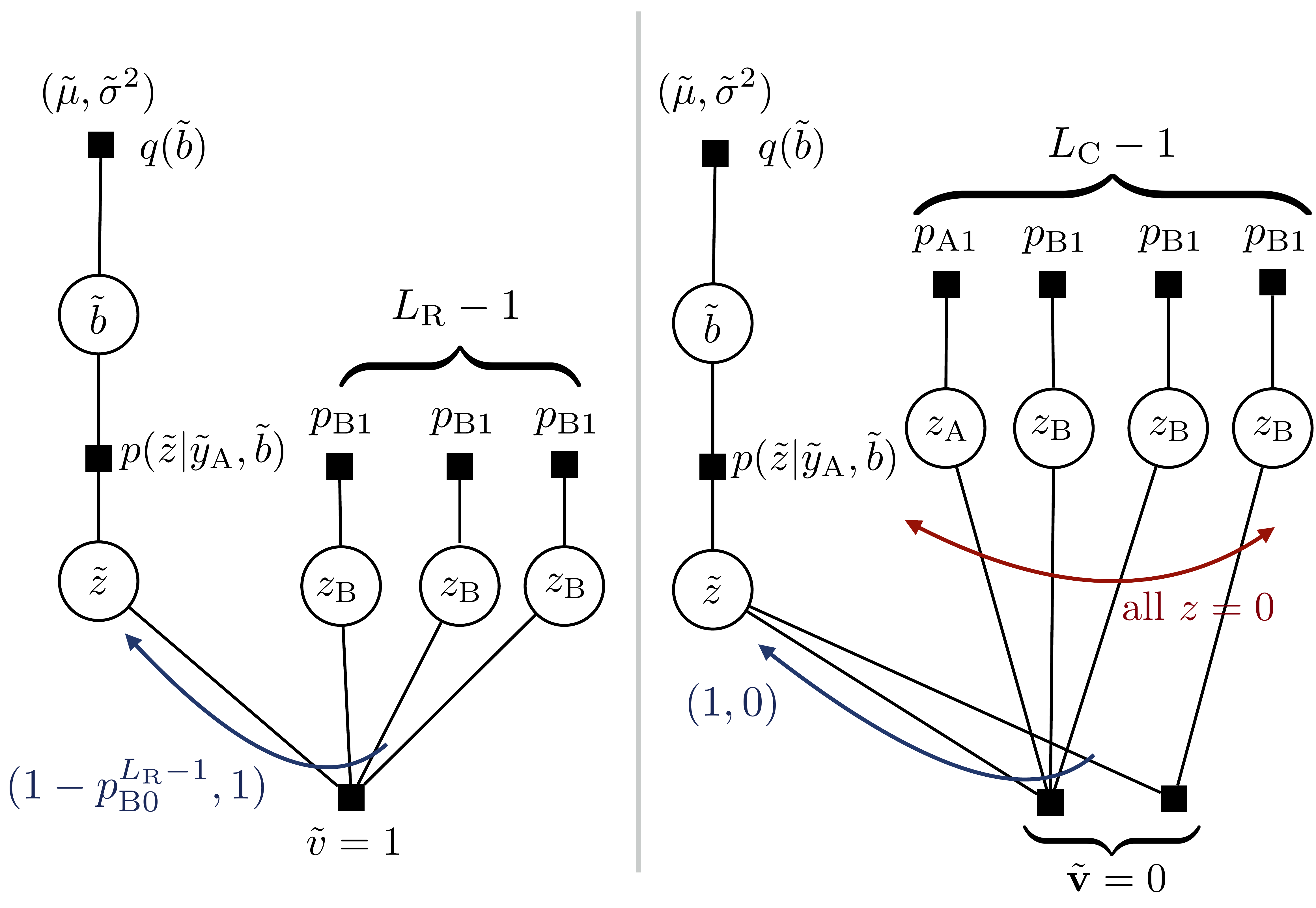}
\caption{On the left, factor graph for the case of congestion event signal observed from the users. The information flowing upwards to the variable node $b$ of the mean field link affects the variance of the posterior belief. On the right, worst case scenario for observing no congestion signals from the users of one network link. We request in this case that all the links connected to the mean field through all of its users to be feasible.}
\label{fig:mean_field}
\end{figure}

The second case corresponds to the event where the mean field link does not trigger a congestion event. In this scenario the belief variance can be reduced only if none of the users employing the link observes a congestion event, see Subsection~\ref{sec:prob_settings}. In this case we can consider as worst case scenario the condition where a user observes no congestion only when none of the employed links is congested. The probability for an uncongested mean field link to have no users with $\tilde{v}=1$ is then equal to:
\begin{equation}
p_{\text{A}0}^{\alpha({L}_C-1)}p_{\text{B}0}^{(1-\alpha)({L}_C-1)},
\end{equation}
where ${L}_C$ is the average number of links that share at least one user with the mean field link, see Fig.~\ref{fig:mean_field}. In mathematical terms  ${L}_C$ is the average number of non-zero entries for the rows of the matrix $\v{A}\v{A}^\tran$. In this case the new variance is equal to the variance of the following distribution:
\begin{equation}
p'(\tilde{b}) = p(\tilde{z}=0|\tilde{y}_A,\tilde{b})q(\tilde{b})/Z_{\text{norm}}.
\label{eq:post_var_0_valid}
\end{equation}

\vspace{0.25cm}

Finally, we need to consider the same analysis when the mean field link belongs to class B. It is easy to see that in this case due to the shape of the likelihood function and the assumption $p_{\text{B}1}\simeq 0$, the  belief variance would basically remain constant even if we could observe the hidden variable $\tilde{z}$. As a result we assume a worst case scenario where the expected posterior variance for the links of class B remains constant.

Combining the two scenarios we can write down an approximate expression for the expected future variance of the mean field model:
\begin{equation}
\begin{aligned}
\mathbb{E}[\tilde{\sigma}^{2\prime}]=  & (1-\alpha) \tilde{\sigma}^2  +\\
			 &\alpha \bigg( p_{\text{A}1} \Big((1-\alpha)^{{L}_R-1}(\tilde{\sigma}^{2\prime}_\text{A1} - \tilde{\sigma}^2) + \tilde{\sigma}^2\Big) +\\
			 &  p_{\text{A}0} \Big( p_{A0}^{\alpha({L}_C-1)}p_{B0}^{(1-\alpha)({L}_C-1)}(\tilde{\sigma}^{2\prime}_\text{A0} - \tilde{\sigma}^2) + \tilde{\sigma}^2  \Big)\bigg),
\label{eq:post_var_total}
\end{aligned}
\end{equation}
where $\tilde{\sigma}^{2\prime}_\text{A1}$ and $\tilde{\sigma}^{2\prime}_\text{A0}$ correspond to the variance of the distribution in Eq.~\eqref{eq:post_var_1_valid} and Eq.~\eqref{eq:post_var_0_valid} respectively.
In Eq.~\eqref{eq:post_var_total} we consider that the only scenarios where the belief variance changes is when a link belongs to class A and the aforementioned worst case condition  are observed. In all the other scenarios the belief variance does not change.
Note that, since we lack of a generative model for $\v{v}$, the conducted analysis does not aim at producing an accurate model for the expected posterior variance, but rather at modeling some situations where we expect to have a considerable variance reduction of the belief on $\tilde{b}$. 

The developed model can now be used in order to minimize the mean field objective function of Eq.~\eqref{eq:MPC_like_MF}.
As independent variables of our model we consider $\alpha$, $p_{\text{A}1}$ and $p_{\text{B}1}$, which are all defined over the interval $[0,1]$. $\alpha$ here represents the ratio of links that belong to class A. Note that since $p(z=1|{y} = \hat{b})$ is monotonically increasing, it can be inverted, and we can find a map from $p_{\text{A}1}$ to $\hat{b}_{\text{A}}$, which is denoted by  $\hat{b}_{\text{A}}(p_{\text{A}1})$. The final problem then becomes:
\begin{equation}
\begin{aligned}
\underset{p_{\text{A}1},p_{\text{B}1},\alpha}{\text{minimize}}\ &\alpha(\hat{b}_{\text{A}}(p_{\text{A}1})-\tilde{\mu})^2 + (1-\alpha)(\hat{b}_{\text{B}}(p_{\text{B}1})-\tilde{\mu})^2 + \tilde{\sigma}^{2} \\ 
& + \frac{\gamma}{1-\gamma}\mathbb{E}[\tilde{\sigma}^{2\prime}](p_{\text{A}1},p_{\text{B}1},\alpha).
\label{eq:mean_field_opt}
\end{aligned}
\end{equation}
All the terms in Eq.~\eqref{eq:mean_field_opt} are well defined and since the optimization problem is non-convex, but it involves only three variables, we can simply solve it by performing a grid search.

\begin{figure}
\centering
\includegraphics[width=1.\columnwidth]{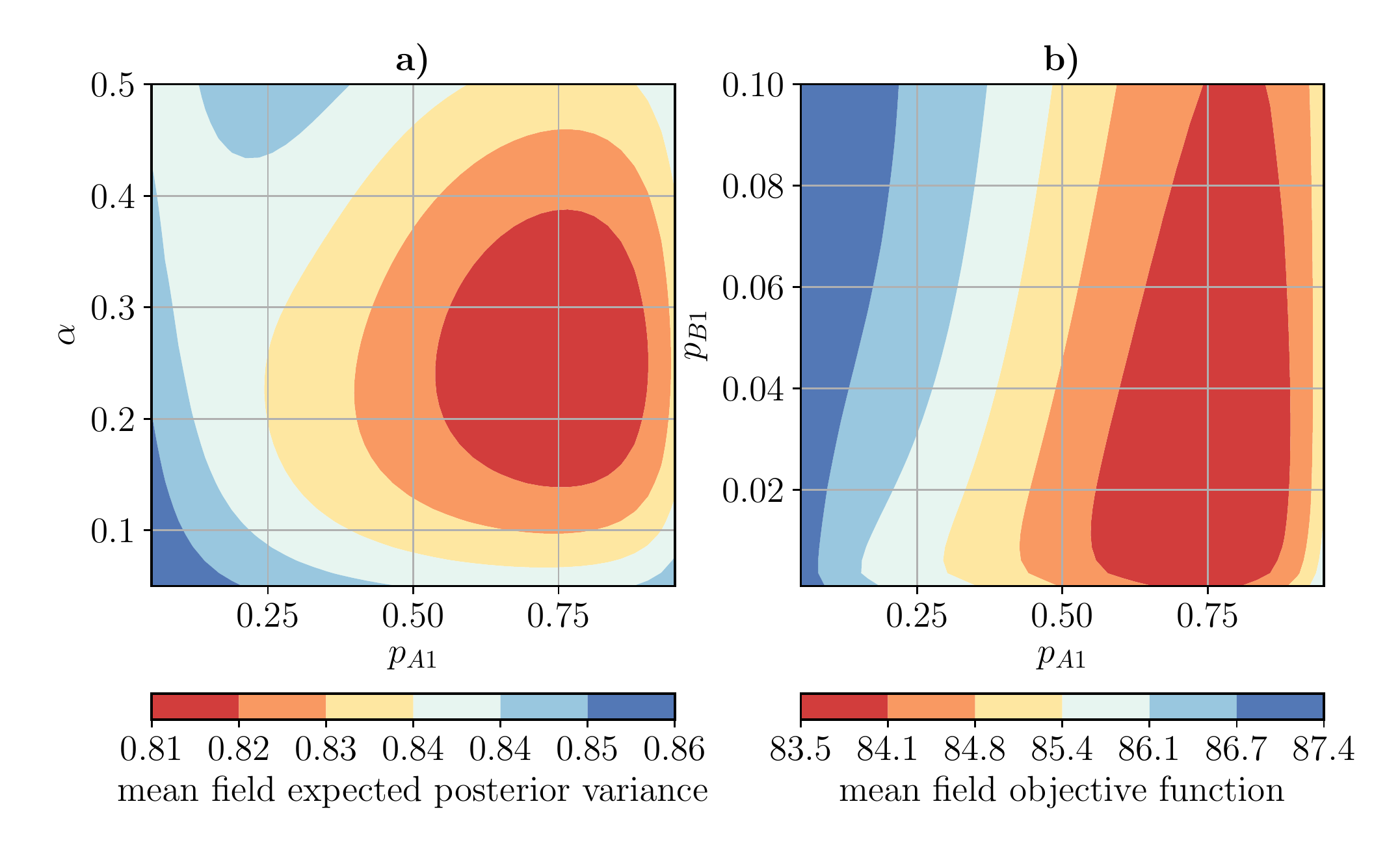}
\caption{\textcolor{navy}{\textbf{a}) Numerical evaluation of the posterior expected variance of the mean field model defined in Eq.~\eqref{eq:post_var_total}. The figure shows how this quantity changes as a function of $p_{\text{A}1}$ and $\alpha$ with a fixed $p_{\text{B}1}=0.01$. \textbf{b}) Numerical evaluation of the objective function of the mean field model defined in Eq.~\eqref{eq:mean_field_opt}. The figure shows how this quantity varies as a function of  $p_{\text{A}1}$ and $p_{\text{B}1}$ ($\gamma=0.99$). For both figures, the other involved parameters have been set to ${L}_R=4$, ${L}_C=20$, and ${\tilde{\sigma}}^2=0.85$, respectively.}}
\label{fig:exp_var}
\end{figure}

In Fig.~\ref{fig:exp_var} we show how the expected posterior variance and the value of the objective function vary as a function of the  optimization variables.
More specifically we show the approximate expected posterior variance as a function of $\alpha$ and $p_{\text{A}1}$. Empirically we observed that the minimizer is usually located at $\alpha\simeq 1/{L}_R$. This is intuitively reasonable, as, in order to reduce the posterior variance in the case of link congestion, we need to have routes that are composed by one link of class $A$ and all the other links of class B. Having one link every  $L_R$ that belongs to class A increases the chances of observing such routes.
The minimizer with respect to $p_{\text{A}1}$ has instead a more complex dependency with respect to the other variables and parameters of the model, however, we observed empirically that it is usually located between $0.5$ and $0.8$ making rather uncertain the possibility of observing a congestion. The $p_{\text{B}1}$ parameter instead minimizes the future variance when is set to 0. In fact, in order to increase the chances to reduce the variance of the links of class A $p_{\text{B}1}$ should be as low as possible. However, a low $p_{\text{B}1}$ increases the deviation from the mean belief and consequently the immediate risk for the links of class B. The optimal learning formulation allows to easily take care of this tradeoff. 
To this extent we can observe in Fig.~\ref{fig:exp_var} that the minimizer of the objective function with respect to $p_{\text{B}1}$ is usually located around $0.01-0.05$. 
Note that the optimal value of $p_{\text{B}1}$ is consistent with the assumption $p_{\text{B}1}\simeq 0$ made  in the analysis conducted above. If this condition was not true at the optimal point, then we could not have used the approximate mean field variance prediction developed in this section.
Finally, we noticed that as long as  the sigmoid function of Eq.~\eqref{eq:z_def} is sharp with respect to the mean field belief $q(\tilde{b})$, i.e., large variance of $q(\tilde{b})$, then the optimal parameters $(p_{\text{A}1},p_{\text{B}1},\alpha)$ are not sensitive to the value of  $\tilde{\sigma}$. As a result, as long as we are uncertain about the capacity values, we are not forced to re-optimize the mean field parameters every time the belief changes.

In the next section we summarize the complete algorithm proposed in this work and we discuss the key points of its implementation.


\section{Overview of the Complete Algorithm}\label{sec:full_alg}
The complete set of operations required to run our algorithm is reported in Alg.~\ref{alg1}. The overall system is composed by two groups of processes, namely the $N$ user and the $M$ link processes (each responsible for an individual network link). First, each user $n$ requests the prices $\lambda_m$ of the employed links to the different resource processes and computes the sending rate according to Eq.~\eqref{eq:primal_update} (lines 3-4). The users then forward the rate $x^t_n$ and the feasibility signal $v_n^t$ to all the process of the employed links (line 5).
At this point the resource processes  compute independently the link rate $y_m^t$ simply by  adding up all the users rates and update the price using Eq.~\eqref{eq:dual_update} (lines 7-8).
The next operation consists in updating the dataset $\{\mathcal{D}^t\}$ with the new observed datapoint (line 9). In order to limit the size of the dataset we subsample the points that are included, and we set a maximum number of stored points (we discard the older points when the maximum size is reached). In this way we can reduce the number of operation required by the EP algorithm. A more detailed description of the updating operation can be found in Appendix~\ref{appA}.

When the inner loop has converged, we recompute the value of the constraint vector $\hat{\v{b}}$ (line 10).
Note that we can establish the convergence of the inner loop algorithm by monitoring the variation of the vectors $\vg{\lambda}^t$ and $\v{y}^t$. Alternatively we can simply update the vector $\hat{\v{b}}$ on a fixed time basis.
The next operation corresponds to running the EP algorithm in a distributed way among the $M$ link processes using the data available in the dataset (line 11). 
Once the belief is updated we use the results from the mean field analysis to set $\hat{\v{b}}$. In order to execute this step we need to know the optimal values for $(\alpha,p_{\text{A}1},p_{\text{B}1})$. This operation can either be computed by each link process independently, or it can be computed by one process and forwarded to the all the link processes. Since we consider the network topology and the sigmoid function characterizing the probability of a link congestion to not change in time,  we can compute the optimal values in advance using the information on the network topology and the initial prior on the network links, then hardcode the optimal values of $(\alpha,p_{\text{A}1},p_{\text{B}1})$ in the link processes. As shown in the results section this simple strategy is sufficient to provide satisfying performance of the proposed algorithm.
Since we know that links belonging to class A are the ones that are likely to reduce their variance, instead to assign randomly at each step $t$ $\alpha M$ links to class A and $(1-\alpha) M$ to class B, we assign to class A the $\alpha M$ links with the largest variance (line 15-16).
Before doing this operation we set the variance of the links that were underutilized in the previous time step to zero (lines 12-14). The intuition is that if links are underutilized, then it is useless to reduce the uncertainty on their capacity value as they are associated to loose constraints, and the optimal rate allocation does not depend on them.
Finally, after the class assignment each link process, given its current belief computes the value of $\hat{b}_m$ that matches the probability to trigger a congestion equal to the value of its class.
Note that the class assignment operation can be done in a fully distributed way by using a consensus algorithm. More details on its implementation can be found in Appendix~\ref{appB}. 

The operation, listed in Alg.~\ref{alg1} are executed continuously. The belief on $\v{b}$ is expected after some steps to converge, shrinking the probability density around some value of $\v{b}$, that should match the true value of the link capacities.

\begin{algorithm}[t]                     
\caption{Complete algorithm}          
\label{alg1}   
\begin{algorithmic}[1]
\Loop 
	\For{each user $n$}
		\State {collect $\lambda^t_m \forall m \in n$}
		\State {compute and apply rate $x^{t}(\v{a}_n^\tran \vg{\lambda}^{t} )$}
		\State {Forward $v_n^t$ and $x_n^t \forall m \in n$}
	\EndFor
	
	\For{each link $m$}
		\State ${\lambda_m}^{t+1} = \max ( 0,{\lambda_m}^{t} + \epsilon({y}_m^t - \hat{{b}}_m^t) )$
		\State $\hat{b}^{t+1}_m = \hat{b}_m^t$
		\State update dataset $\{\mathcal{D}^t\}$
	\EndFor
	
		
		\If {inner loop converged}
			\State compute $q^t(\v{b})$ using EP
			
			\For{each link $m$}
				\If {${\lambda}_m^t=0$}
					\State ${\sigma_m^2} = 0$
				\EndIf
			\EndFor
			
			\State assign to class A $M \alpha$ links with largest $\sigma^2_m$
			\State assign to class B other links

			\For{each link $m$}
				\State find $\hat{b}_m\ :\ p(z'=1|\hat{b}_m)=p^\text{class}_m$
			\EndFor
			\State $\hat{b}^{t+1}_m = \hat{b}_m$
		\EndIf
\EndLoop
\end{algorithmic}
\end{algorithm}

\textcolor{black}{
We now briefly discuss at a high-level the complexity of the algorithm in terms of communication and storage cost.
The inner loop of the proposed system corresponds to a classical NUM algorithm: at each step $t$ each user communicates with the resource processes of the links composing the route to collect the link prices and forward the sending rate and congestion signal. Considering an average route made of $L_R$ links, this operation involves a transmission of $\mathcal{O}(NL_R)$ messages.
The execution of the EP algorithm is the most expensive part in terms of communication and storage requirements of the entire algorithm. Each link process $m$ has to store, for each episode $t$ in the dataset, the route of the users who employed link $m$ and detected a congestion event $v_n=1$ (plus the link rate $y_m$). Therefore, if we denote by $N_{\text{link}}$ the average number of users per link, the storage cost is $\mathcal{O}(N_{\text{link}}L_R)$ for each element in the dataset and for each link process. 
$N_{\text{link}}L_R$ basically corresponds to the average number of factor nodes $\psi^1_n$ connected to each variable node $z_m^t$ in the factor graph of Fig.~\ref{fig:factor_graph}. This means that, in terms of communication requirements, one complete update of the incoming messages for all the variable nodes in the factor graph requires an exchange of $\mathcal{O}(N_{\text{link}}L_R)$ messages for each link process and for each episode in the dataset. 
EP on a factor graph with loops has to be executed multiple times on the entire dataset before convergence. Unfortunately, it is not easy to quantify the number of iterations required by the EP algorithm to converge. In our implementation we stop the EP update after ten iterations, since empirically we have observed that they are usually sufficient to provide a good belief estimate.
Considering fixed the amount of refinements required by the EP algorithm, the overall cost, in terms of communication and storage, at time $t$ is $\mathcal{O}(N_{\text{link}}L_Rt)$ for each process $M$.
The complexity grows linearly with time because the dataset is obviously growing. In order to avoid this we limit the size of the dataset to the last $S_\mathcal{D}$ observations. The key point in the complexity analysis is the dependency of $L_R$ with respect to $N_{\text{link}}$. The answer to this question however depends on the network topology. For instance, if the network  belongs to the class of small-world networks, and user routes correspond to the shortest routes between the source and destination nodes, then $L_R$ grows logarithmically with respect to the number of total network nodes, making the overall complexity of the algorithm more tractable.}
\textcolor{black}{Finally, the mean field optimization is basically independent of the numbers of links and users. The class assignment task uses a simple method derived from classical consensus average algorithms~\cite{consensus}, and simply involves exchange of local messages among the $M$ link processes.}
This concludes the discussion on the implementation of the proposed algorithm, in the next section we show and discuss the performance of the proposed method.

\section{Simulation Results}\label{sec:results}

In order to test the proposed method we generate some random networks with different numbers of links and users. We assign to each user a log-shaped utility function $u_n(x)=w_n \log(x)$, where  $w_n$ is a random positive parameter, and a random route connecting two (non adjacent) nodes of the network.
After the routing matrix is set, we sample the prior distribution of the link capacities in order to set their true value. We consider that each link has a lognormal distribution with mean and standard deviation equal to $(\mu^0_m,\sigma^0_m)$. We set $\mu^0_m = 2.7N_m$ where $N_m$ is equal to the number of users using the link and $\sigma^0_m=0.27N_m$. 
Regarding the parameters of the sigmoid function of Eq.~\eqref{eq:z_def} we fix $\rho=0.95$ and $\kappa=5((1-\rho)b_m)^{-1}$. In this case we have that when  $y_m=0.95b_m$ there is a 0.5 probability for link $m$ to trigger a congestion event, whereas when $y_m=b_m$ the probability goes up to about $0.99$.
We optimize the values of $(p_{\text{A}1},p_{\text{B}1},\alpha)$ at the beginning of each simulation using the $L_\text{R}$ and $L_\text{C}$ of the network in use, the average mean and variance of the prior distribution of the different links, and the parameters $\kappa$ and $\rho$ defined above.
In order to generate the observation vector $\v{v}$ at each step $t$, we first generate the link congestion vector $\v{z}$ using Eq.~\eqref{eq:z_def} and then we randomly pick a value of $\v{v}$ among the ones that are consistent with the model specified in Subsection~\ref{sec:prob_settings}.

\begin{figure*}[h]
	\centering
	\includegraphics[width=1.\textwidth]{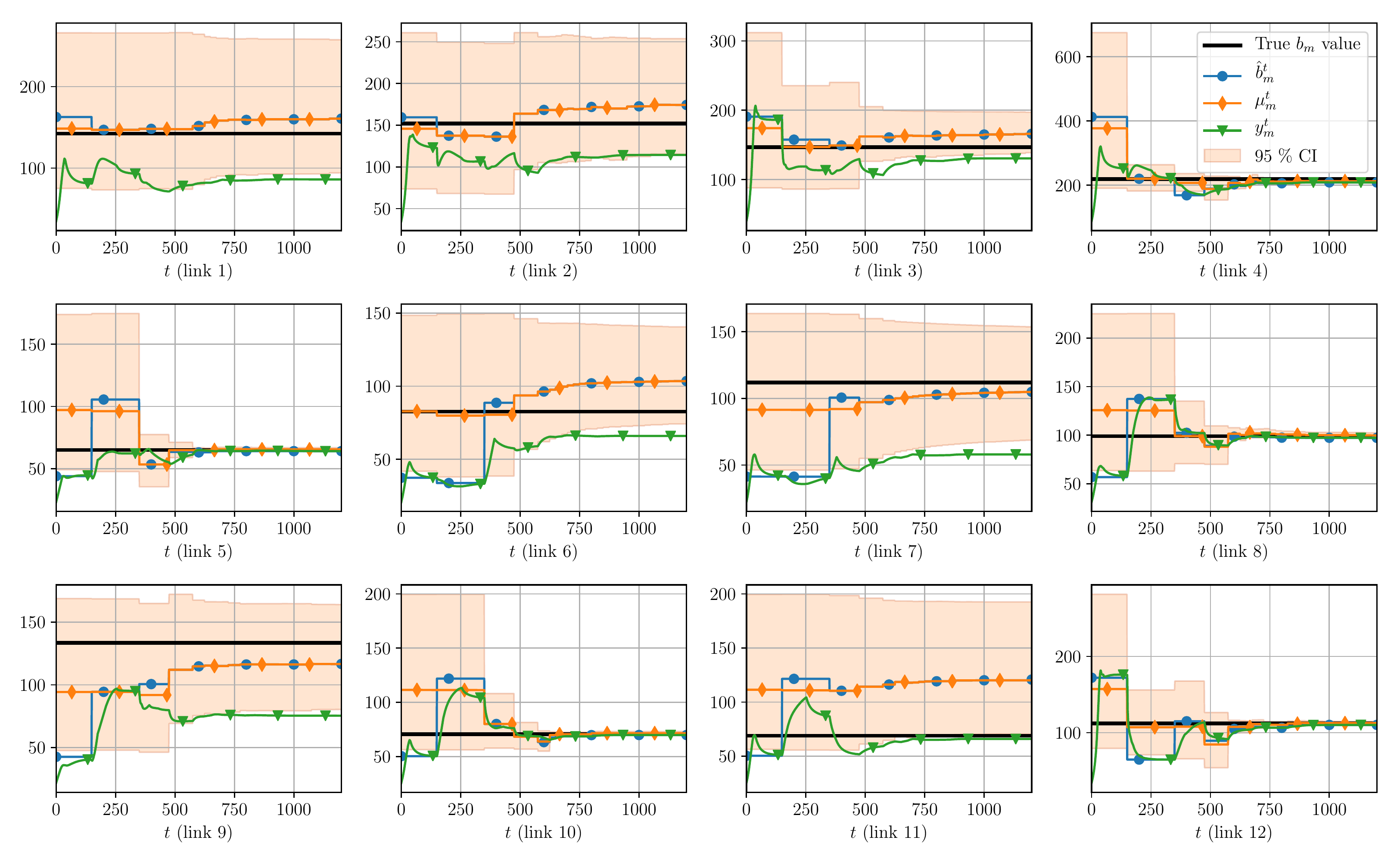}
	\caption{Evolution of the most important quantities employed by the proposed algorithm for all the $M$ links. Each subplot shows: the true $b_m$ value, the evolution of the mean value of the lognormal belief of $b_m$, the $95$\% Confidence Interval (CI) of the belief, the value of the decisions $\hat{\v{b}}^t$, and finally, the users sending rate through link $m$ $y_m$.}
	\label{fig:all_links}
\end{figure*}

In the first test we run the algorithm on a network with $12$ links  and $200$ users and  a discount factor of $\gamma=0.99$, Fig.~\ref{fig:all_links} shows the evolution of some quantities involved in the algorithm operation.
The plots show the evolution of the mean value of the lognormal belief of $q^t(b_m)$, the $95$\% Confidence Interval (CI) of the belief, the value of $\hat{\v{b}}^t$, and finally, the sum of the users sending rates for link $m$, $y_m$.
The selection of $\hat{\v{b}}^t$ is consistent with the results of the mean field approximation. For example, between $t=0$ and $t=200$, the value of $\hat{b}_1$ (which has a large variance)  is set to a value that is slightly larger than the mean value, whereas for link $5$ and $6$ $\hat{b}$ is set to a value below the mean of their belief. 
We can verify our assumption about the approximation $\hat{\v{b}}=\v{y}$. As it can be seen, the requested link rate $y_m$ tends to converge to the value of  the parameter $\hat{{b}}_m$. When it does not reach the value $\hat{{b}}_m$ it is because the link capacity is too large and the link is underutilized at equilibrium. The assumption $\hat{\v{b}}=\v{y}$ holds for the remaining active links.
Concerning the underutilized links, we can observe that, since the rate $y_m$  reached by the inner loop is much lower than the mean value of the capacity belief, the uncertainty of the constraint cannot be reduced and it remains rather high. However, in this case, there is no need to reduce the variance of these capacities as they do not affect the optimal rate allocation. For the active links, the algorithm continues to sample at values of $y_m$ close to the mean value of the belief. As a result, the variance sequentially shrinks and the mean value approaches the true value of $b_m$.

\begin{figure}
	\centering
	\includegraphics[width=1.\columnwidth]{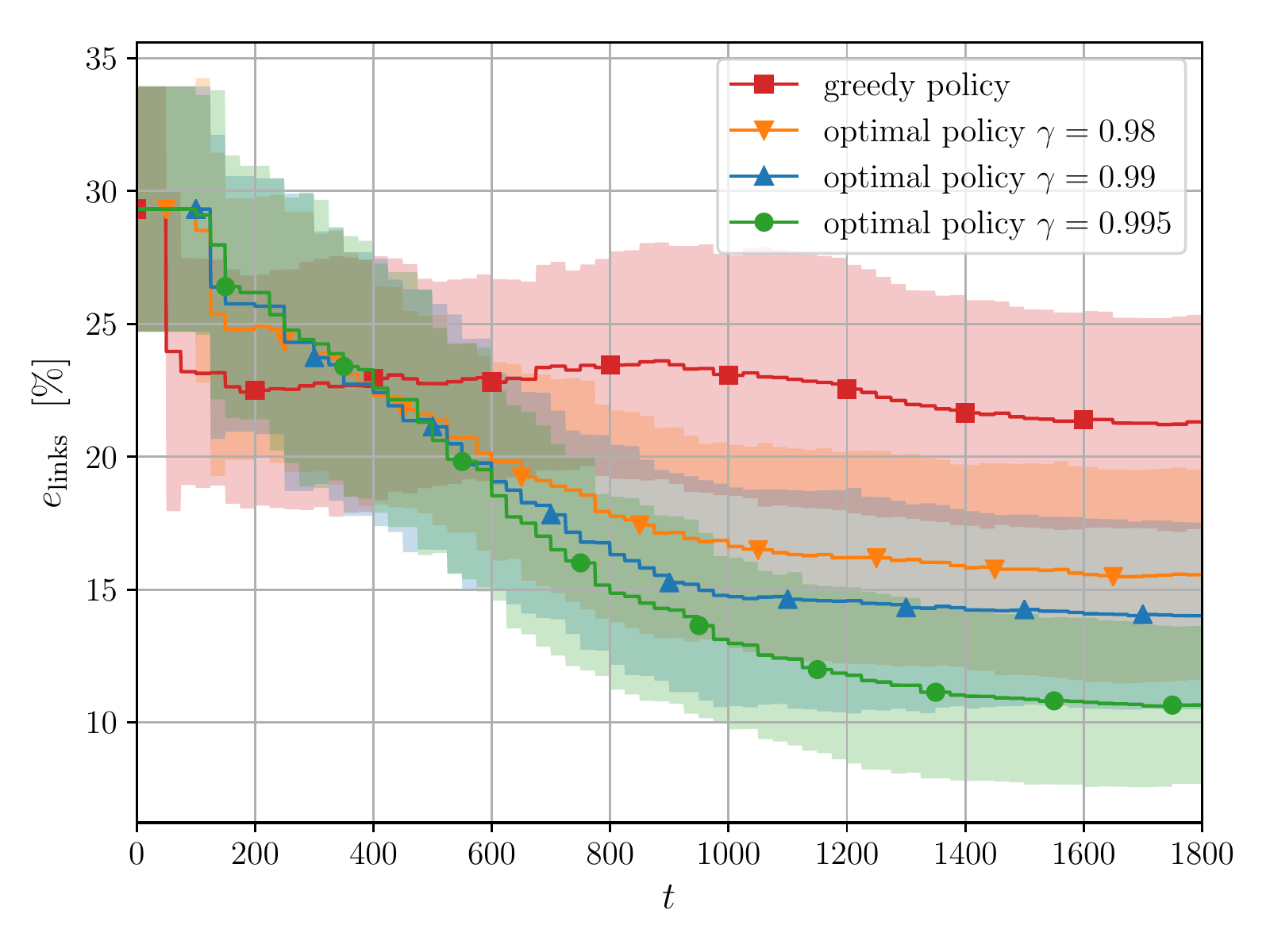}
	\caption{Evolution of the mean absolute percentage error between ${\mu}_m^t$ and the true vector ${b}_m^{\text{true}}$. The plot shows the average and standard deviation of the metric among ten different runs.}
	\label{fig:b_diff}
\end{figure}

\begin{figure}
	\centering
	\includegraphics[width=1.\columnwidth]{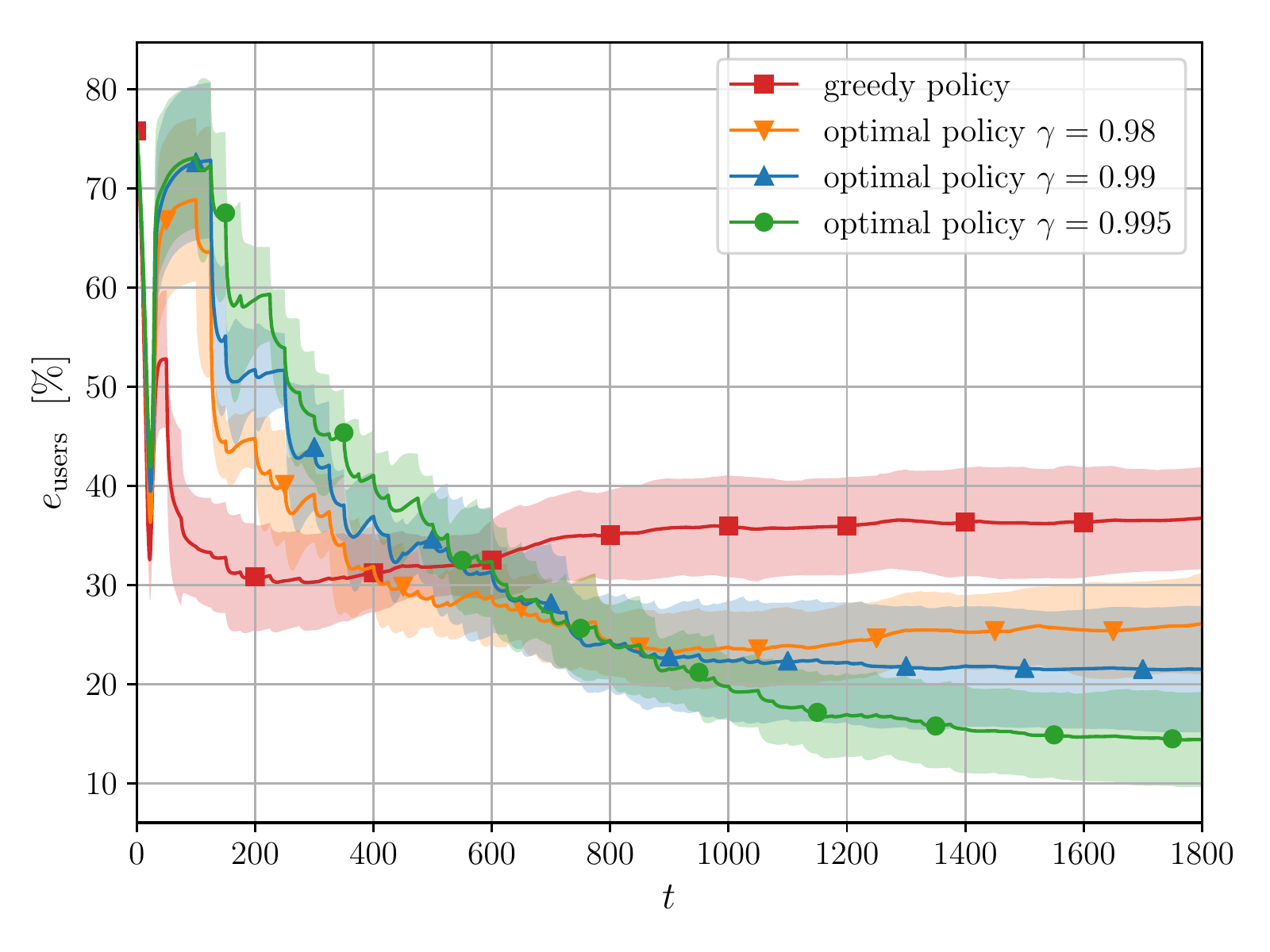}
	\caption{Evolution of the mean absolute percentage error between the rates ${x}_m^t$ used and the optimal rates ${x}_n^{\star}$. The plot shows the average and standard deviation of the metric among ten different runs.}
	\label{fig:g_diff}
\end{figure}

\textcolor{navy}{In order to evaluate the performance of the rate allocation algorithm we compare it to a greedy algorithm that simply minimizes the immediate risk (i.e., $\gamma=0$) and matches the value of $\hat{\v{b}}$ with the mean value of the current available belief. The greedy algorithm does not seek for values of $\hat{\v{b}}$ that are expected to reduce the future belief uncertainty, and resembles a passive method that does not perform any active learning.}
For our algorithm we set the values of the future risk discount $\gamma$ to $0.98$, $0.99$ and $0.995$.
We use a topology with $48$ links and $800$ users.
We draw ten different sets of samples of $\v{b}$ from the prior distribution and ten different sets of values for the parameters $w$ of the utility functions. We then run the proposed algorithm with different $\gamma$ values and the greedy method on the same ten different random settings.
In Fig.~\ref{fig:b_diff} we plot the evolution of the average value over the different runs, and the standard deviation, of the mean absolute percentage error for the capacity values. More specifically the metric used corresponds to:
\begin{equation}
e_\text{links} = 100\% \sum_m \frac{|{{\mu_m}} - {b}_m^{\mathrm{true}}|}{M{b}_m^{\mathrm{true}}}
\end{equation}
where ${\mu_m}$ represent the mean of the current belief.
The greedy algorithm is actually able to reduce the parameters uncertainty by a larger value in the very first stages but then it struggles to further reduce the uncertainty which remains constant  around $22\%$ throughout the entire simulation. Running the system with $\gamma \simeq 1$ instead leads to lower values for the mean absolute percentage error, between $10\%$ and $17\%$ for the different values of $\gamma$. The error decreases slowly at the early stages, however the reduction is more persistent and eventually achieves much lower mean error. Moreover we can see that larger values of $\gamma$  lead to lower long term error.
Due to the existence of underutilized links we cannot expect the error to decrease to zero. In fact for these links the average error is expected to remain large  since the belief variance does not decrease.

Another metric that can be used to evaluate the algorithm is the evolution of the mean absolute percentage error of the users' sending rate with respect to the optimal ones:
\begin{equation}
e_\text{users} = 100\% \sum_n \frac{|x_n - {x}_n^{\star}|}{N{x}_n^{\star}}\
\end{equation}
This quantity is not sensitive to potential underutilized links and directly measures how the users' rates are close to the optimal ones. The evolution of this metric is shown in Fig.~\ref{fig:g_diff}. As for the previous figure, we show the mean and standard deviation among ten different runs.
In order to reduce the future risk our algorithm sacrifices the immediate one, achieving a larger error in the first steps with respect to the greedy strategy. However, in the final steps, when the capacity beliefs are more accurate, the users' rates tend to be closer to the optimal ones and our method, for $\gamma=0.995$, settles around $e_\text{users} \simeq 15\%$. The greedy algorithm instead is not able to significantly reduce the error for the future steps, and settles to an average percentage error of $37\%$.

\begin{figure}
	\centering
	\includegraphics[width=1.\columnwidth]{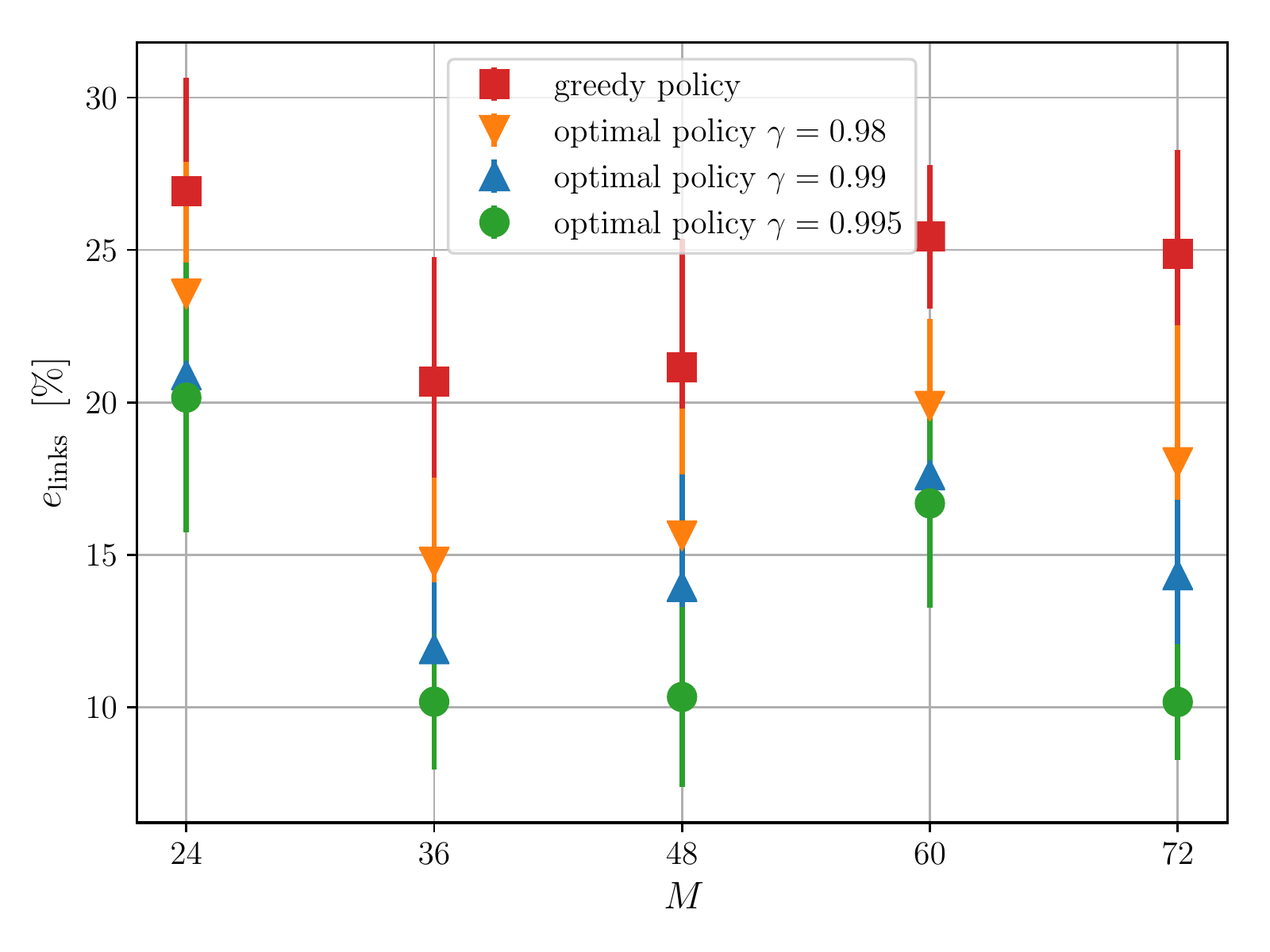}
	\caption{Mean absolute percentage error between ${\mu}_m^t$ and the true vector ${b}_m^{\text{true}}$ for networks with a different number of links $M$. The plot shows the average and standard deviation of the metric among ten different runs.}
	\label{fig:topo_b}
\end{figure}

\begin{figure}
	\centering
	\includegraphics[width=1.\columnwidth]{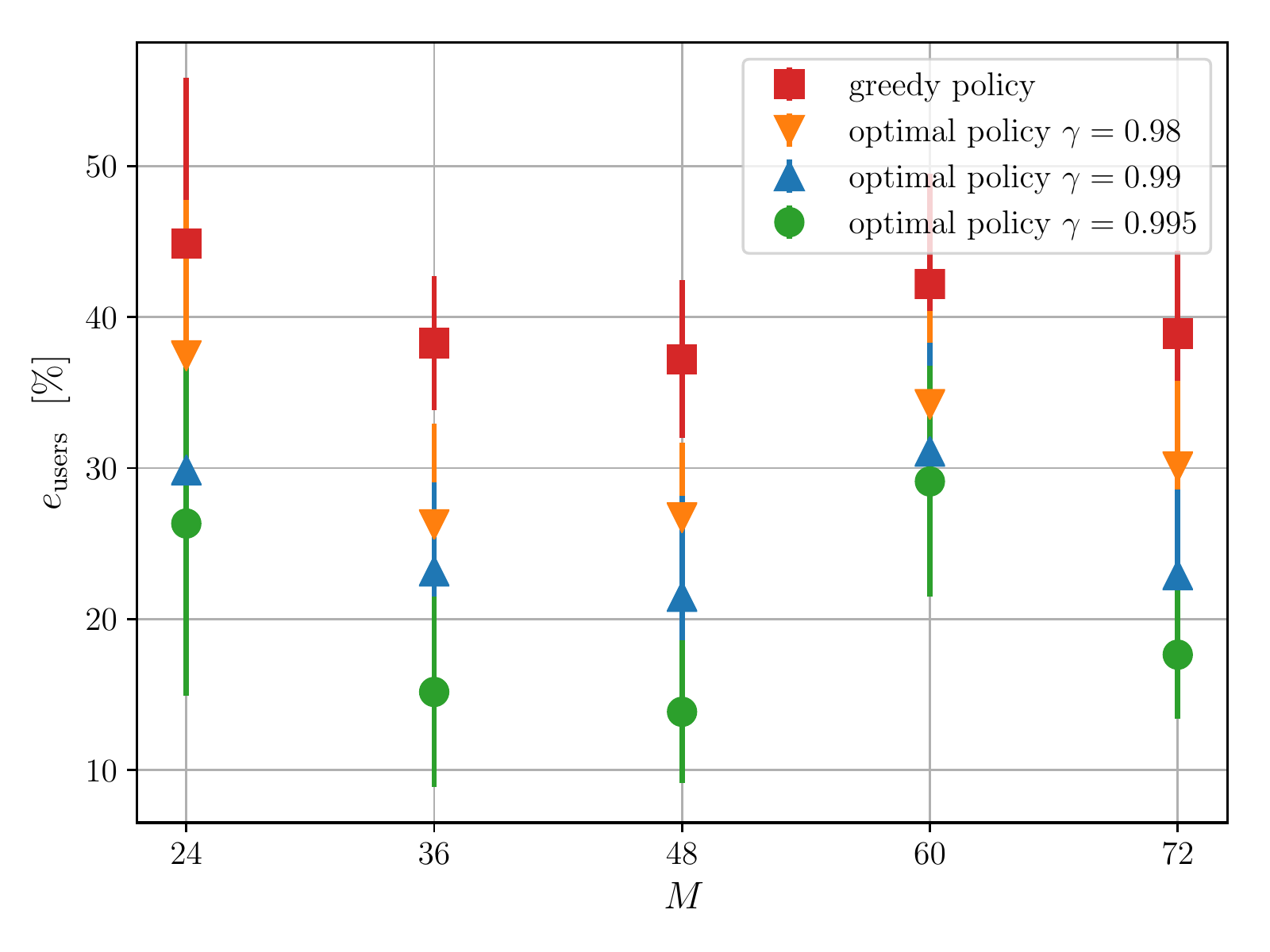}
	\caption{Mean absolute percentage error between the rates ${x}_m^t$ used and the optimal rates ${x}_n^{\star}$ for networks with a different number of links $M$. The plot shows the average and standard deviation of the metric among ten different runs.}
	\label{fig:topo_g}
\end{figure}

We further evaluate the performance of the algorithm for different network sizes. We generate five different networks, with $M=[24,36,48,60,72]$ links and $N=[400,600,800,1000,1200]$ users. For each network we draw ten different sets of values for  $\v{b}$ and ${w}$. Though the network size differs for the different simulations we generate the users routes in order to have an average route length of $4$ links for each topology. We run the simulation for the greedy policy and for $\gamma$ equal to $0.98$, $0.99$ and $0.995$.
As for the previous tests we compute the mean absolute percentage error of the capacity estimates and the users' rates. We then plot the mean and standard deviation for the ten different runs after $1800$ iterations. The results are depicted in Fig.~\ref{fig:topo_b} and Fig.~\ref{fig:topo_g}.
The best performance is achieved with $\gamma=0.995$, with $e_\text{users}$ between $15-30\%$. the greedy strategy achieves the worst performance for all the different network sizes, with about $38-45\%$ of error with respect to the optimal user rate.
The results show that the performance of the algorithm  is not strongly correlated with the network size. The performance discrepancies among the different sizes are likely due to the  different random realizations of the network topology.

\begin{figure}
	\centering
	\includegraphics[width=1.\columnwidth]{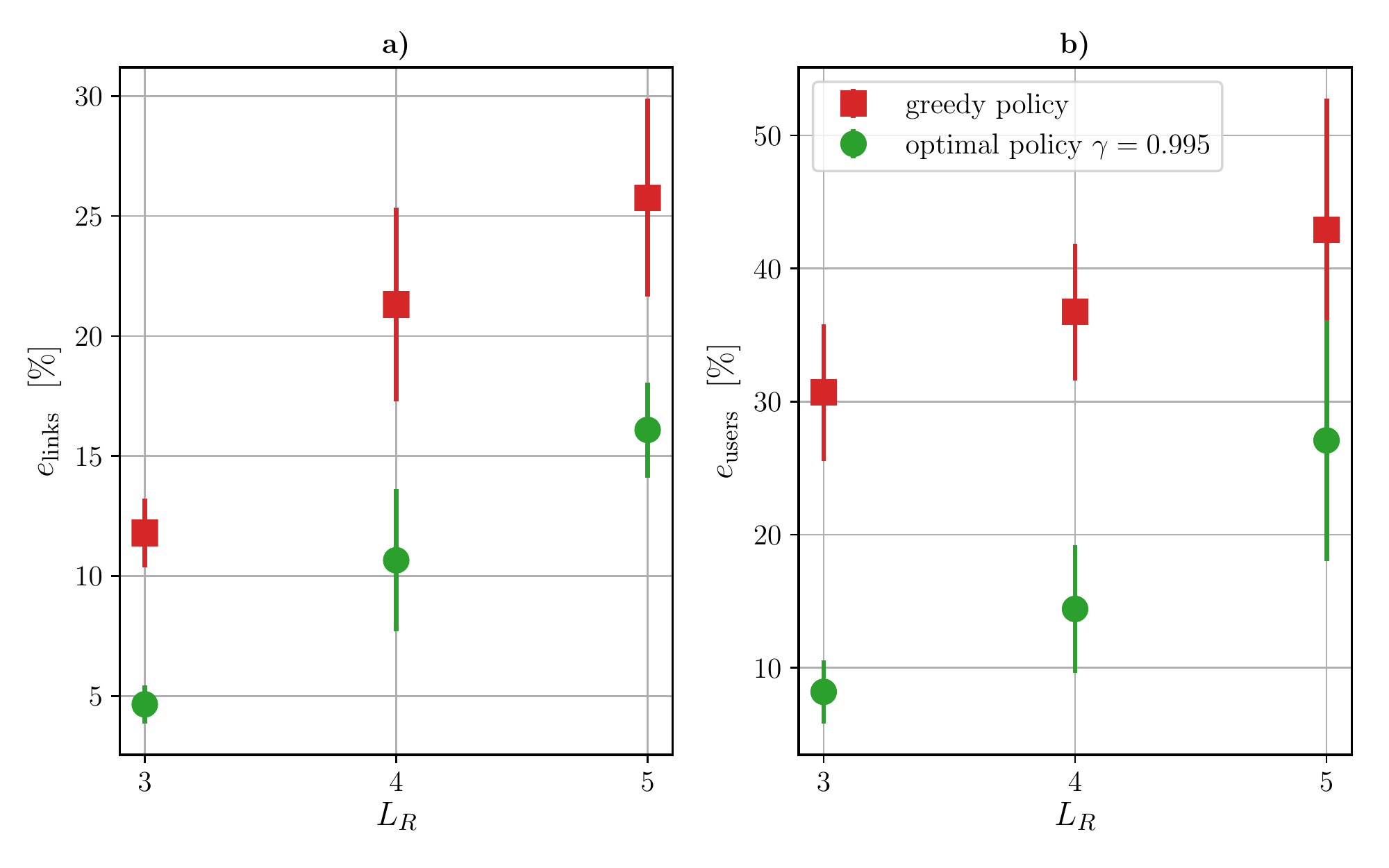}
	\caption{\textbf{a}) Mean absolute percentage error between ${\mu}_m^t$ and the true vector ${b}_m^{\text{true}}$ for different average route length $L_R$. \textbf{b}) Mean absolute percentage error between the rates ${x}_m^t$ used and the optimal rates ${x}_n^{\star}$ for different average route length $L_R$ The plot shows the average and standard deviation of the metric among ten different runs..}
	\label{fig:route_all}
\end{figure}

In the final simulation we investigate how the average length of the user routes affects the performance of the proposed method. We use a network topology with $M=48$ and we generate ten different realizations of the user populations for different values of the average route length, with $L_R=[3,4,5]$. We then compute the mean and standard deviation of the same metrics used in the previous experiments for the different scenarios. The results are shown in Fig.~\ref{fig:route_all}.
The results show that, when the average route length increases the system performance decrease for both the greedy policy and the proposed foresighted method. However the proposed method always outperforms the greedy policy. This is somewhat expected since longer routes make it more challenging for the inference method to correctly estimate the value of the latent variables $\v{z}$. As a consequence, the value of the link capacities is also harder to infer.

\section{Conclusions}\label{sec:conc}
In this work we consider a specific instance of the NUM problem where the amount of the network resources is unknown and the private congestion signals of the users cannot be used to directly achieve the optimal rate allocation. The congestion signals, however, can be combined to infer the amount of available resources. We design a distributed overlay rate allocation method where the users communicate with other helper processes, one for each network link, in order to achieve the optimal rate allocation. The proposed solution method consists in decomposing the original problem into two subproblems: one subproblem corresponds to the classical NUM problem, the second subproblem instead corresponds to the design of an adaptive controller for the classical NUM algorithm that infers the amount of available resources. Using an optimal learning formulation we are able to balance the exploration versus exploitation tradeoff that arises in the design of the adaptive controller and guarantee good performance of the system in the long run. As shown by the conducted evaluation, such performance cannot be attained using a greedy strategy.

We  believe that the analyzed problem and proposed framework, though being in its early stages, could be of great interest for the future computer networks. Consider for instance the Internet: the challenges and requirements of the platforms using the network evolve faster than the underlying infrastructure. In this scenario if the new applications want to optimize the data transmission they might have to coordinate the users using only information from the endpoints, since these are the network parts that are accessible and can be updated more easily. The new applications might then need to infer what are the actual global resources available. Though being extremely challenging and complex, building an overlay allocation method that infers the available resources and adapts to different network conditions, might be the only viable solution if the lower layers, and the infrastructure of the communication network, cannot be changed.

\section*{Acknowledgment}
This work has been supported by the Swiss National Science Foundation under grant CHISTERA FNS 20CH21 151569.

\appendices  


\section{Expectation Propagation Implementation}\label{appA}

The implemented EP algorithm on the considered factor graph of Fig.~\ref{fig:factor_graph} is reported in Alg.~\ref{alg0}. We denote by $\mu$ and $\nu$ the incoming and outgoing messages respectively, of the different variable nodes. These messages are possibly unnormalized probability density (mass) functions defined over the domain of the variable node. Their subscripts denote the variable node and factor node of the message\footnote{To make the notation more clear the messages to and from the factor node $p(z_m^t|b_m,y_m)$ are indexed by the variable nodes  $b_m$ and $z^t_m$.}. 
Since the factor graph, depicted in Fig.~\ref{fig:factor_graph}, has loops, the incoming messages of the variable nodes, ${z}^t_m$ and $b_m$, need to be iteratively refined (line 1) till they converge to their final value. All the messages are initialized using non-informative distribution (lognormal distribution with infinite variance for the $b_m$ variables and uniform distribution for the binary variables $z_m^t$).
At each iteration the algorithm updates the incoming messages $\mu_{b_m z_m^t}$ parallelizing the operations over $M$ different processes.
Each process $m$ computes the incoming message to the $z_m^t$ from the $b_m$ using the messages from the $t'\neq t$ observations and the prior information (line 4).
Then we iterate over the individual factors of $\Psi_{\v{v}^t}()$ to update the incoming messages of the latent variables.
Each process $m$ computes the outgoing message from the $z_m^t$ node to the factor node $\psi_n^1()$ associated to any observation $n$ with $v_n^t=1$ (lines 5-7).
As next step, each process collects all the outgoing messages $\nu_{z^t_l \psi_n^1}$  for $l\neq m $ and $a_{ln}=1$ and computes the incoming message $\mu_{z^t_m \psi_n^1}$ (lines 8-9):
\begin{equation}
\mu_{z^t_m \psi_n^1}(r_m) = \sum_{\substack{l\neq m\\ r_l \in \{0,1\}}} \psi_n^1(\v{r})\prod_{\substack{l\neq m\\ a_{ln}=1}} \nu_{z^t_{l} \psi_n^1}(r_l),
\label{eq:incomingz}
\end{equation}
where we have emphasized the fact that the messages $\mu$ and $\nu$ are actually functions.
A na\"ive computation of this quantity is  rather expensive in terms of required operations. However, since $\psi_n^1(\v{r})$ is equal to one when at least one $r_m$ variable employed by user $n$ is equal to one and zero otherwise, we can compute $\psi_n^1(\v{r})$ as the complementary event of having all the $r_m$ variables equal to zero. We therefore obtain: 
\begin{equation}
\begin{aligned}
\mu_{z^t_m \psi_n^1}(0) &= \prod_{\substack{l\neq m\\ a_{ln}=1}} (\nu_{z^t_{l} \psi_n^1}(0) + \nu_{z^t_{l} \psi_n^1}(1)) - \prod_{\substack{l\neq m\\ a_{ln}=1}} \nu_{z^t_{l} \psi_n^1}(0)\\
\mu_{z^t_m \psi_n^1}(1) &= \prod_{\substack{l\neq m\\ a_{ln}=1}} (\nu_{z^t_{l} \psi_n^1}(0) + \nu_{z^t_{l} \psi_n^1}(1))
\label{eq:incomingz_simp}
\end{aligned}
\end{equation}
if we consider to normalize the outgoing messages $\nu_{v^t_{l} z^t_n}$ we obtain the simpler form:
\begin{equation}
\begin{aligned}
\mu_{z^t_m \psi_n^1}(0) &= 1 - \prod_{\substack{l\neq m\\ a_{ln}=1}} \nu_{z^t_{l} \psi_n^1}(0)\\
\mu_{z^t_m \psi_n^1}(1) &= 1.
\label{eq:incomingz_simp_norm}
\end{aligned}
\end{equation}
This is a simple operation that consists in the product of the outgoing messages associated with the event $z_m^t=0$ for the observation $v_n^t=1$. The factors $\psi_m^0()$ can easily be handled locally by each link process. 
Once all the incoming messages from all $\v{v}^t$ factors are updated, the incoming message $\mu_{b_m z_m}^t$ can be computed. The  EP algorithm then aims at minimizing the following divergence in order to obtain the belief for $b_m$:
\begin{equation}
q = \underset{q'}{\text{arg min}}\ \  \text{KL}(\mu_{b_m z^t_m}\prod_{t'\neq t}\tilde{\mu}_{b_m z^{t'}_m}||q'),
\label{eq:KLEP}
\end{equation}
where $q(b_m)$ represents the lognormal distribution characterizing the belief on $b_m$. In order to minimize Eq.~\eqref{eq:KLEP} we simply need to match the sufficient statistics between the two distributions of the KL divergence. In order to match the sufficient statistics we need to perform integration over the first argument of the KL divergence of Eq.~\eqref{eq:KLEP}. As the integral is a simple one dimensional integral of a well behaved function any common numerical integration method can be used to achieve this task.
The message $\tilde{\mu}_{b_m z^{t}_m}$ which corresponds to the lognormal approximation of ${\mu}_{b_m z^{t}_m}$ is then set to:
\begin{equation}
\tilde{\mu}_{b_m z^{t}_m} = \frac{q}{\prod_{t'\neq t}\tilde{\mu}_{b_m z^{t'}_m}}.
\label{eq:factorapprox}
\end{equation}
Note that all the terms in Eq.~\eqref{eq:factorapprox} are lognormal distributions, which belong to the exponential family, therefore the above operation simply consists in algebraical manipulation of the distribution parameters. The EP algorithm allows also to add a damping factor ($\epsilon_{\text{EP}}$) in the update of the incoming messages $\mu$ (and $\tilde{\mu}$) see~\cite{MPDM}. This modification prevents the messages from changing drastically with respect to their previous value; the equilibrium points do not change but it can improve the convergence of the algorithm.
The above steps are executed for each datapoint in the dataset and for $I_\text{MAX}^{\text{EP}}$ iterations. At this point, the execution stops and the lognormal distribution $q(\v{b})$ defined by the most updated values of the messages $\tilde{\mu}_{b_m z^{t}_m}$ approximates the true posterior $p(\v{b}|\{\mathcal{D}^t\})$.

\begin{algorithm}[t]                     
\caption{Implementation of the EP algorithm.}          
\label{alg0}   
\begin{algorithmic}[1]

\For{$i<I_\text{MAX}^{\text{EP}}$}
	\For{ each observation to process $t$}
		\For{each link $m$}
			\State update $\mu_{z^t_m b_m}$
		\EndFor
		\For{each $n$ with ${v}_n^t=1$}
			\For{each link $m$ with $a_{mn}=1$}
				\State compute $\nu_{z^t_m v^t_n}$
				\State collect $\nu_{z^t_l v^t_n}$ $\forall l\neq m,\ a_{ln}=1$
				\State update $\mu_{z^t_m v^t_n}$
			\EndFor
			\State compute $\mu_{b_m z^t_m}$
			\State $q = \underset{q'}{\text{arg min}}\ \  \text{KL}(\mu_{b_m z^t_m}\prod_{t'\neq t}\tilde{\mu}_{b_m z^{t'}_m}||q')$
			\State $\tilde{\mu}_{b_m z^t_m} = \frac{q}{\prod_{t'\neq t}\tilde{\mu}_{b_m z^{t'}_m}}$

		\EndFor
	\EndFor
\EndFor			
\end{algorithmic}
\end{algorithm}

\begin{figure}
	\centering
	\includegraphics[width=\columnwidth]{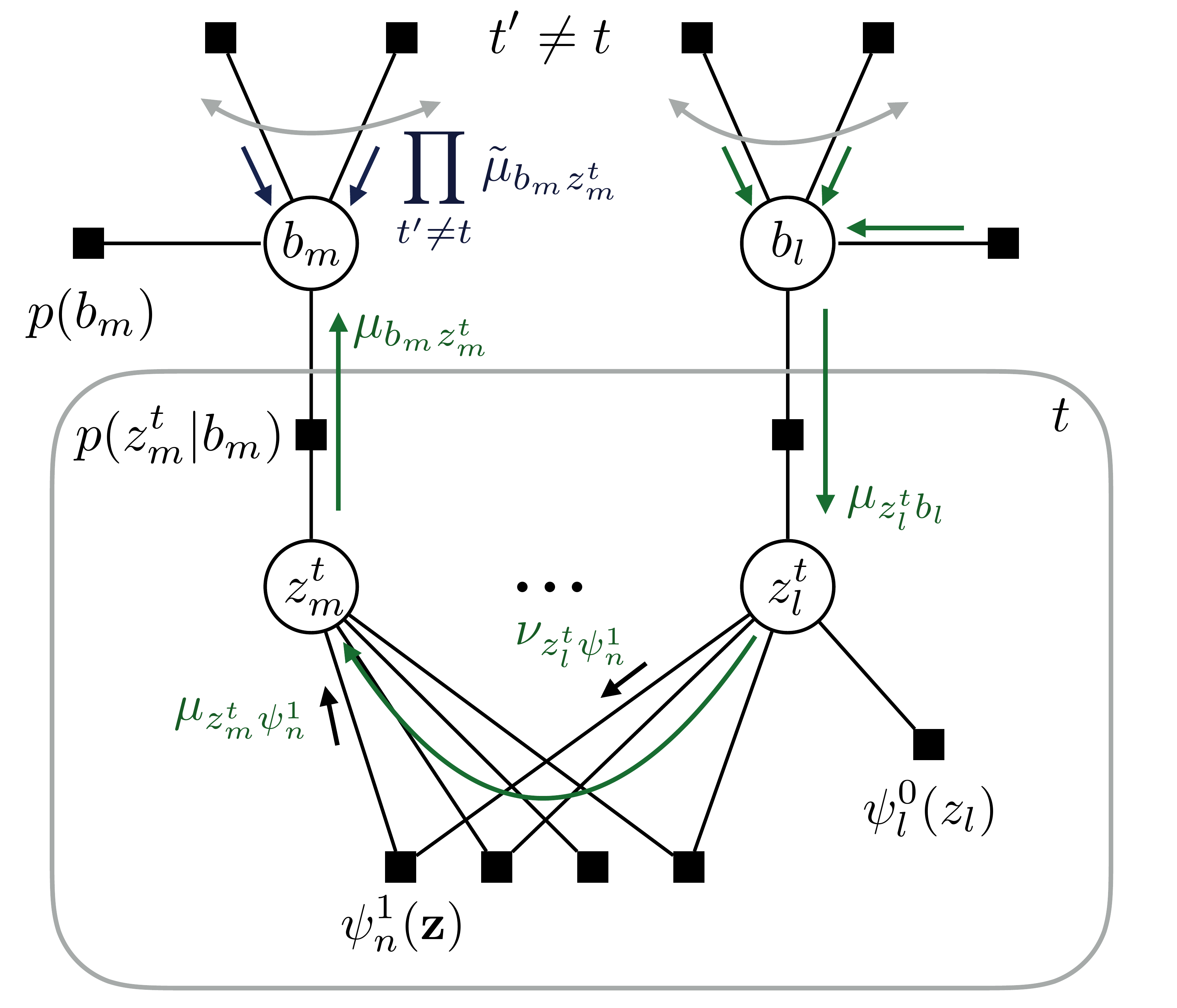}
	\caption{Factor graph of one single observation $t$ showing the messages between the variables nodes exchanges during the EP algorithm execution.}
	\label{fig:factor_graph}
\end{figure}

\section{Detailed Algorithm Implementation}\label{appB}

\begin{algorithm}[t]                     
\caption{Complete algorithm}          
\label{alg1}   
\begin{algorithmic}[1]
\Loop 
	\For{each user $n$}
		\State {collect $\lambda^t_m \forall m \in n$}
		\State {compute and apply rate $x^{t}(\v{a}_n^\tran \vg{\lambda}^{t} )$}
		\State {Forward $v_n^t$ and $x_n^t \forall m \in n$}
	\EndFor
	
	\For{each link $m$}
		\State $y_m^t = \v{a}_m^\tran \v{x}^t$
		\State ${\lambda_m}^{t+1} = \max ( 0,{\lambda_m}^{t} + \epsilon({y}_m^t - \hat{{b}}_m^t) )$
		\State $\hat{b}^{t+1}_m = \hat{b}_m^t$
	\EndFor
	
	\If {$\text{modulo}(t,T_S)=0$}
		\For{each link $m$}
			\State {update dataset $\{\mathcal{D}^t\}$ with point $({y}_m^t,{v}_n^t)$} 
		\EndFor
		
		\If {$|\lambda_m^{t+1} - \lambda_m^{t} |\leq V_\lambda \ \forall m$}
			\State compute $q^t(\v{b})$ using EP
			
			\For{each link $m$}
				\If {$\vg{\lambda}^t=0$}
					\State $\vg{\sigma^2} = 0$
				\EndIf
			\EndFor
			
			\State Initialize $\hat{\v{r}}^i=0$,  ${\v{r}^i}=0$, $i=0$
			\While{$|{h}^i_m - {h}^{i-1}_m|\leq V_h \ \forall m$}
				\For{each link $m$}
					\State $r^i_m = \underset{{r}'\in \{0,1\}}{\text{arg max}}\ {r}'({\sigma}_m^2-{h}^i_m)$
					\State $\hat{r}^{i+1}_m = 1 - \v{l}_m^\tran \hat{\v{r}}^{i} + \Delta r_m^i$
					\State $h^{i+1}_m =\left(1 - \v{l}_m^\tran \v{h}^{i} + \delta_h(\hat{r}_m^i - \alpha)\right)^+$ 
					\State $p^\text{class}_m = r^i_mp_{\text{A}1} + (1-r^i_m)p_{\text{B}1}$
				\EndFor
			\EndWhile 

			\For{each link $m$}
				\State find $\hat{b}_m\ :\ p(z'=1|\hat{b}_m)=p^\text{class}_m$
			\EndFor
			\State $\hat{b}^{t+1}_m = \hat{b}_m$
		\EndIf
	\EndIf
\EndLoop
\end{algorithmic}
\end{algorithm}

The detailed complete set of operations required to run our algorithm is reported in Alg.~\ref{alg1}. The overall system is composed by two groups of processes, namely the $N$ users and the $M$ link processes (each responsible for an individual network link).
As described in the main manuscript lines (3-8) correspond to the execution of the classical NUM algorithm. The next operations are instead specific to our system.

The next step corresponds to updating the dataset $\{\mathcal{D}^t\}$ with observations $\{\v{y}^t,\v{v}^t\}$. Though our design optimizes the choice of $\hat{\v{b}}^t$ by taking into account the expected posterior variance when observing exclusively the point $\{\v{y}^t,\v{v}^t\}$ after the inner loop has converged, we could as well use the information collected during the inner loop dynamics. However, points that share similar link rates $\v{y}^t$ provide redundant information to the inference method, moreover a higher number of points increases also the computation cost of the running the EP algorithm. In order to have a tradeoff between the two cases we adopt a simple heuristic that consists in adding one observations to the dataset every $T_S$ time steps (lines 10-12). Moreover in order to limit the amount of computations of the EP algorithm we limit the maximum size of the dataset to $S_\mathcal{D}$. When the maximum size is reached we discard the older points to make space for the new ones. The most recent values of the approximated incoming messages  $\tilde{\mu}_{b_m z^{t}_m}$ for the  discarded points are however embedded in the prior belief.
When the inner loop has converged (line 13) we execute the distributed EP algorithm, see Alg.~\ref{alg0}, on our graphical model using the current dataset and obtain the new fitted distribution $q^t(\v{b})$ (line 14).

\begin{figure*}[t]
\normalsize
\begin{subequations}
\begin{empheq}[left={\empheqlbrace\,}]{align}
&\;\;\v{r}^{i}= \underset{\v{r}'\in \{0,1\}^M}{\text{arg max}}\ \v{r}'^\tran(\vg{\sigma}^2-\v{h}^{i})\\ 
&\begin{bmatrix}
\hat{\v{r}}^{i+1} \\  
\v{h}^{i+1}
\end{bmatrix}
=\left[
\begin{array}{c|c}
\v{I} - \v{L} & \v{0} \\
\hline
\delta^i_h \v{I} & \v{I} -\v{L}
\end{array}
\right]
\begin{bmatrix}
\hat{\v{r}}^i \\  
\v{h}^{i} 
\end{bmatrix}
- \left[
\begin{array}{c|c}
\v{I}  & \v{0} \\
\hline
\v{0} & -\delta^i_h \v{1}
\end{array}
\right]\left[ \Delta\v{r}^i | \alpha \right] \\[4pt]
&\;\;\v{h}^{i+1}= (\v{h}^{i+1})^+
\end{empheq}
\label{eq:complete_update}
\end{subequations}
\hrulefill
\end{figure*}

At this point we need to compute the new value of $\hat{\v{b}}$ to use in the next iterations.
In order to set the value of $\hat{\v{b}}$ according to the mean field approximation, we need a rule to assign each link either to class A or to class B. Our solution consists in assigning to class $A$ the links with the larger variance. According to our model, the links of class A are actually the ones expected to reduce more significantly their belief variance. 
Before assigning the links to the different classes it is however beneficial to modify the current variances $\sigma^2_m$ for the links that are actually underutilized as  their values do not affect the rate allocation.
We set the variance $\sigma^2_m$ of these links to zero, basically forcing them to belong to class B (line 15-17).
We design a distributed algorithm for the selection of the $\alpha M$ largest  links with the largest variance $\sigma^2_m$ among the $M$ link processes. We can formulate the problem as an optimization problem:
\begin{equation}
\begin{aligned}
\underset{\v{r}}{\text{maximize}} \ \ &\ \v{r}^\tran\vg{\sigma}^2\\
\text{subject to} \ \ &\   \v{1}^\tran \v{r} \leq \alpha M \\
\ \ &\ \v{r} \in \{0,1\}^M,
\end{aligned}
\label{eq:selection_orig}
\end{equation}
where $\v{r}$ is a binary vector representing the class membership of link $m$: if $r_m=0$ then the link $m$ belongs to class B and to class A otherwise. For simplicity we assume here that $\alpha M \in \mathbb{N}$ and that the solution to problem in Eq.~\eqref{eq:selection_orig} is unique. If these conditions hold we can solve exactly the above problem using a dual method, which allows to decompose the solution method among the $M$ processes.  We consider the Lagrange relaxation of Eq.~\eqref{eq:selection_orig}:
\begin{equation}
\underset{\v{r} \in \{0,1\}^M}{\text{maximize}} \ \v{r}^\tran\vg{\sigma}^2 - h \left( \v{1}^\tran \v{r} - \alpha M \right),
\label{eq:lagrange}
\end{equation}
where $h$ denotes the Lagrange multiplier of the inequality constraint. Given the value of the dual variable $h$ each process can compute independently the value of $r_m$. This operation is extremely simple, basically $r_m=1$ if $\sigma^2_m>h$ and zero otherwise. In order to solve the dual problem the variable $h$ can be iteratively updated using a gradient descent method:
\begin{equation}
h' = \big( h + \delta_h \left(\hat{r} - \alpha \right)\big)^+,
\label{eq:del_dual_update}
\end{equation}
where $\hat{r}$ represents the mean value of the elements of $\v{r}$ and $\delta_h$ is a positive parameter that controls the step length. Unfortunately this update operation requires a central entity which knows the value of $\hat{r}$ and forwards the dual variable $h$ to all the $M$ link processes. Ideally we prefer to have an algorithm that is completely distributed, and does not require a central entity, therefore we modify the iteration in the following way. We first create $M$ copies of the dual variable $h$ and $M$ estimates of the mean value $\hat{r}$, one for each link process. Each process then, iteratively executes Eq.~\eqref{eq:del_dual_update} using its local copies and runs in parallel a consensus algorithm on both variables~\cite{consensus}. In our case we need to solve a dynamic average consensus problem, which means that the nodes have to agree on the average value of a $M$ dimensional input signal that is varying over time (in our case the $M$-dimensional signals are the dual variables $\v{h}$ and the estimates $\hat{\v{r}}$).
In order to run a consensus algorithm among the $M$ link processes we define a connected undirected graph $\mathcal{G}=(\mathcal{V},\mathcal{E})$, where the nodes $\mathcal{V}$ correspond to the $M$ processes and the edges $\mathcal{E}$ denote the pairs of processes that exchange messages in the class assignment method. We define by $\v{L}$ the normalized Laplacian matrix of the graph $\mathcal{G}$.
Many works focused on how to optimize communication among the agents in order to speed up the convergence of average consensus algorithms (see~\cite{optimgraph}) in this work, however, we simply assume that the $M$ links processes define a random connected communication network. 


The iterative steps of the class assignment algorithm are summarized in Eq.~\eqref{eq:complete_update},
where $\Delta\v{r}=\v{r}^i-\v{r}^{i-1}$ denotes the variation of vector $\v{r}$ with respect to previous iteration, and $\v{I}$ denotes the identity matrix. Each iteration is composed by three steps.
First each process selects the optimal value of $r_m$ according to the local copy of the dual variable $h^i_m$.
The variables $\hat{\v{r}}$ are then updated  according to the following dynamics $\hat{\v{r}}^{i+1} =(\v{I} - \v{L})\hat{\v{r}}^{i} + \Delta\v{r}^{i}$. Considering null initial conditions, we have that, under a steady input $\v{r}$ the vector $\hat{\v{r}}$ converges to $\v{1}\sum_m r_m/M$ (similar dynamic consensus methods have been proposed in~\cite{dyncontconsensus,dyndiscconsensus}). 
Note that, at each step ,each link process has to communicate with the other processes that correspond to its neighbors on the graph $\mathcal{G}$ in order to collect the values of $\hat{r}^{i}_l$.
At the same time, the dual variables $\v{h}$ evolve according to $\v{h}^{i+1} =(\v{I} - \v{L})\v{h}^{i} + \delta (\hat{\v{r}}^{i} - \v{I}\alpha)$. The operations are basically the same as before, except taht each link process integrates the difference between the target ratio  $\alpha$ and the local estimate of the mean value of $\v{r}$. The use of the Laplacian matrix $\v{L}$ in this case assures that possible discrepancies among the entries of the vector $\v{h}$ are damped, making all the link processes to agree on the same value of the dual variable. Finally, the variables $\v{h}$ are constrained to be non-negative to be consistent with to Eq.~\eqref{eq:del_dual_update}.
Executing iteratively Eq.~\eqref{eq:complete_update} the link processes agree on the class assignment, and the links that have a largest uncertainty are assigned to class A (lines 18-24).
At this point each link process computes the value of $\hat{b}_m$ that makes $p(z_m=1|\hat{b}_m)$ equal to the probability of the link class (lines 25-26).

Finally, all the parameters used in our simulations for Alg.~\ref{alg0} and Alg.~\ref{alg1} are given in Table~\ref{tab:param}.
\renewcommand{\arraystretch}{1.5}
\begin{table}[h]
\centering
\caption{Parameters settings}
\label{tab:param}
\begin{tabular}{|c|c||c|c|}
\hline
Parameter          & Value  & Parameter          & Value  \\ \hline \hline
$\epsilon_m$             &     $0.5/\hat{b}_m$   &$T_S$          &     $25$        \\ \hline
$V_\lambda$          &      $10^{-3}$    &$V_h$          &    $10^{-3}$       \\ \hline
$\delta^i_h$ & $0.1/\sqrt{i}$ & $\gamma$          &       $0.98-0.995$      \\ \hline
$I_\text{MAX}^{\text{EP}}$          &      $5$     & $\epsilon_{\text{EP}}$          &      $0.5$    \\ \hline 
$S_\mathcal{D}$          &      $25$     &          &       \\ \hline 
\end{tabular}
\end{table}

%

%
%
%
%



\bibliographystyle{IEEEtran}
\bibliography{./learningdraft}
\end{document}